\documentclass[twocolumn]{aa}  
\usepackage{verbatim}
\usepackage{color}

\newcommand       \cm         {\,{\rm cm}}
\newcommand       \mum      {\,{\rm \mu m}}
\newcommand       \km         {\,{\rm km}}

\newcommand       \mol          {\,{\rm mol}}
\newcommand     \gtsim  {\lower.5ex\hbox{$\buildrel > \over \sim$}}
\newcommand     \ltsim  {\lower.5ex\hbox{$\buildrel < \over \sim$}}
\newcommand     \simgt  {\lower.5ex\hbox{$\buildrel > \over \sim$}}
\newcommand     \simlt  {\lower.5ex\hbox{$\buildrel < \over \sim$}}
\newcommand       \simali       {{\sim\,}}
\begin{document}

\title{The infrared bands of polycyclic aromatic hydrocarbons
in the 1.6--1.7$\mum$ wavelength region}

   \author{Tao Chen
          \inst{1}\fnmsep\thanks{Email: taochen@kth.se}
          \and 
          Yi Luo\inst{1,2}\fnmsep\thanks{Email: luo@kth.se}
          \and
          Aigen Li
          \inst{2}\fnmsep\thanks{Email: lia@missouri.edu}
          }
          
   \institute{
         School of Engineering Sciences in Chemistry, Biotechnology and Health, Department of Theoretical Chemistry and Biology, Royal Institute of Technology, 10691, Stockholm, Sweden
         \and
         Hefei National Laboratory for Physical Science at the Microscale, Department of Chemical Physics, School of Chemistry and Materials Science, University of Science and Technology of China, Hefei, 230026 Anhui, China
         \and
         Department of Physics and Astronomy, University of Missouri, Columbia, MO 65211, USA
          }

\abstract
  {The 3.3$\mum$ aromatic C--H stretching band of polycyclic aromatic hydrocarbon (PAH) molecules seen in a wide variety of astrophysical regions is often accompanied by a series of weak satellite bands at $\simali$3.4--3.6$\mum$. One of these sources, IRAS~21282+5050, a planetary nebula, also exhibits a weak band at $\simali$1.68$\mum$. While the satellite features at $\simali$3.4--3.6$\mum$ are often attributed to the anharmonicities of PAHs, it is not clear whether overtones or combination bands dominate the 1.68$\mum$ feature.}
     {In this work, we examine the anharmonic spectra of eight PAH molecules, including anthracene, tetracene, pentacene, phenanthrene, chrysene, benz[a]anthracene, pyrene, and perylene, to explore the origin of the infrared bands in the 1.6--1.7$\mum$ waveelngth region.}
     {Density Functional Theory (DFT) in combination with the vibrational second-order perturbation theory (VPT2) is utilized for computing the anharmonic spectra of PAHs. To simulate the vibrational excitation process of PAHs, the Wang-Landau random walk technique is employed.} 
     {All the dominant bands in the 1.6--1.7$\mum$ wavelength range and in the 3.1--3.5$\mum$ C--H stretching region are calculated and tabulated. It is demonstrated that combination bands dominate the 1.6--1.7$\mum$ region, while overtones are rare and weak in this region. We also calculate the intensity ratios of the 3.1--3.5$\mum$ C--H stretching features to the bands in the 1.6--1.7$\mum$ region, $I_{3.1-3.5}/I_{1.6-1.7}$, for both ground and vibrationally excited states. On average, we obtain $\langle I_{3.1-3.5}/I_{1.6-1.7}\rangle \approx 12.6$ and $\langle I_{3.1-3.5}/I_{1.6-1.7}\rangle \approx 17.6$ for PAHs at ground states and at vibrationally excited states, respectively.}
     {}


   \maketitle

\section{Introduction}
The broad infrared (IR) emission bands at 3.3, 6.2, 7.7, 8.6, and 11.3$\mum$ observed in a wide variety of Galactic and extragalactic objects are generally attributed to polycyclic aromatic hydrocarbon (PAH) molecules \citep{leger1984identification, allamandola1989interstellar}. These bands account for $\simali$10--20\% of the total IR emission of the Milky Way and star-forming galaxies \citep{tielens2008}. Besides these five prominent bands, a few minor bands are also widely observed, e.g., the band at $\simali$3.4$\mum$ which is often attributed to the C--H stretch of aliphatic hydrocarbons \citep{pendleton2002organic}, superhydrogenated PAHs \citep{bernstein1996hydrogenated, sandford2013infrared}, or anharmonicity \citep{barker1987anharmonicity}, and in some objects, multiple weaker features at 3.46, 3.51, and 3.56$\mum$ are also observed, with a tendency to decrease in strength with increasing wavelength \citep{geballe1985spectroscopy, jourdain1986new, joblin1996spatial}. In addition, a weak band at $\simali$1.68$\mum$ is seen in IRAS~21282+5050, a planetary nebula \citep{geballe1994detection}.\footnote{%
   Attempts to search for the 1.68$\mum$ feature 
   in the planetary nebula Hb~5 \citep{magazzu1992non},
   the protoplanetary nebula HD~44179, 
   the planetary nebula He~2--177, and 
   the photodissociated region Orion bar
   \citep{siebenmorgen1993search}
   have been made but only upper limits have been placed. 
   }

There is a debate on the origin of the bands in the 1.6--1.7$\mum$ wavelength region. \cite{brenner1992} predicted the intensity ratio of the 1.68$\mum$ band to the 3.4$\mum$ band to be $I_{1.68}/I_{3.4}\approx1/6$, provided that the 3.4$\mum$ feature arises from the $\nu=2\rightarrow1$ transition while the 1.68$\mum$ feature arises from the $\nu=2\rightarrow0$ overtone emission, where $\nu$ is the vibrational quantum number. However, observationally, \cite{siebenmorgen1993search} derived $I_{1.68}/I_{3.4} \simlt 1/48$ and \cite{magazzu1992non} derived $I_{1.68}/I_{3.4} \simlt 1/45$. \cite{duley1994overtone} argued that the 1.68$\mum$ feature could not be attributed to the overtones of the C--H stretch, instead, it could result from combination bands. On the other hand, experimentally, \cite{reddy1982highly} detected a band at $\simali$1.67$\mum$ in the spectra of highly vibrationally excited benzene and attributed it to the overtone transition.

The theoretical anharmonic spectra of PAHs contain detailed information about each vibrational mode \citep{mackie2015anharmonic, mackie2016anharmonic, mackie2018anharmonic}, which ought to provide conclusive evidence regarding the essence of the bands in the 1.6--1.7$\mum$ wavelength region. Anharmonic spectra can be calculated with {\it ab initio} methods, which, in theory, has been well established for decades \citep{clabo1988systematic, allen1990systematic, amos1991anharmonic, maslen1992higher,martin1995anharmonic, assfeld1995degeneracy}. However, due to high computational costs, in practice, the calculation of anharmonic spectra for large molecules, e.g., PAHs, emerges only recently. 

\cite{maltseva2015high, maltseva2016high, maltseva2018high} measured the IR absorption spectra at very low temperature ($\simali$4\,K) by means of the UV-IR ion dip spectroscopy, allowing them to obtain high-resolution IR spectra of PAHs. Nevertheless, such spectra can hardly be interpreted by harmonic models. To interpret the measured spectra, \cite{mackie2015anharmonic, mackie2016anharmonic, mackie2018anharmonic} calculated the ground-state anharmonic IR spectra for linear, non-linear, hydrogenated and methylated PAHs using the second-order vibrational perturbations theory (VPT2). Although for most of the bands their calculated spectra accurately match the high-resolution experimental spectra, \cite{maltseva2015high} noticed that there are several intensive bands still missing in the C--H stretch region of the calculated spectra. \cite{chen2018carrier} accounted for the missing bands through the inclusion of the 1-3 and 2-2 Darling-Dennison resonances in the VPT2 calculations. To model the astronomical environments and interpret the observations, \cite{mackie2018fully} and \cite{chen2018temperature} incorporated vibrational excitations in the calculations of the anharmonic spectra using the Wang-Landau random walk technique. This approach generated reasonable emission spectra in comparison with the experimental spectra measured by the Berkeley Single Photon InfraRed Emission Spectrometer \citep{schlemmer1994, wagner2000peripherally}.

In this work, we compute the ground-state and vibrationally-excited anharmonic spectra of a number of PAH species to explore the origin of the bands in the 1.6--1.7$\mum$ wavelength region and compare the strengths of the bands in this region with that of the C--H stretch in the 3.1--3.5$\mum$ region.

\section{Computational details} 
The anharmonic vibrational spectra are calculated using the density of functional theory (DFT) as implemented in the Gaussian 16 package \citep{frisch2016gaussian}. The functional of B3LYP \citep{becke,lee1988development} in combination with the polarized double-$\zeta$ basis set, N07D \citep{barone2014fully}, are utilized for the calculations. It has been shown that such a combination produces more accurate spectra in comparison with other DFT methods \citep{mackie2018anharmonic, chen2018carrier}. 

The geometry optimizations are performed with a very tight convergence criterion and a very fine integration grid (Int = 200 974) for numerical integrations. For the resonant calculation, the generalized VPT2 (GVPT2) approach \citep{barone2014fully} is used. GVPT2 involves a two-step procedure: first, resonant terms are identified by means of an ad hoc test \citep{martin1995anharmonic} and successively removed, which is called deperturbed VPT2 (DVPT2). In the second step, the discarded terms are reintroduced through a variational treatment. This approach has been recognized to give  accurate results \citep{maslen1992higher,boese2004vibrational,barone2014fully,mackie2015anharmonic, mackie2016anharmonic,martin1995anharmonic}. 

To balance the computational costs and accuracy of anharmonic IR spectra, only 1-2 Fermi resonances, 1-1 and 2-2 Darling-Dennison resonances are taken into account. The maximum resonant thresholds are set to 200 cm$^{-1}$ for both types of resonances, and the minimum thresholds are all set to 0$\cm^{-1}$. Such settings have been evaluated and produced reasonable spectra \citep{chen2018carrier}.

\section{Results and discussion}
\begin{figure}
\includegraphics[width=1.0\columnwidth]{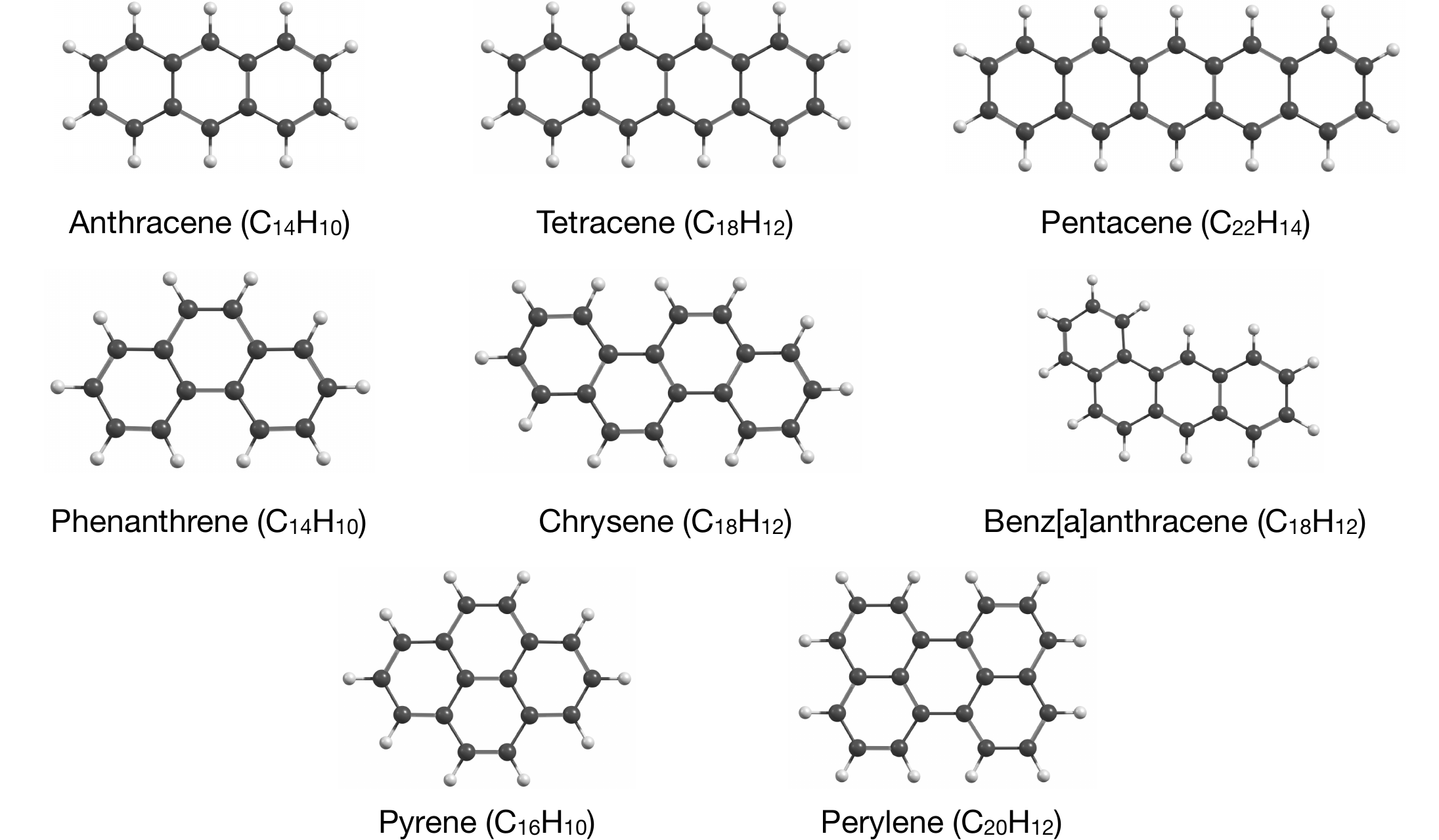}
\caption{The Molecules studied in this work. Anthracene, tetracene and pentacene are linear PAHs. Phenanthrene, chrysene and benz[a]anthracene are non--linear PAHs. Pyrene and perylene are compact PAHs.}
\label{fig:molecules}
\end{figure}

The molecules studied in this work are shown in Figure~\ref{fig:molecules}. They can roughly be divided to three groups: (i) linear PAHs including anthracene (C$_{14}$H$_{10}$), tetracene (C$_{18}$H$_{12}$) and pentacene (C$_{22}$H$_{14}$); (ii) non-linear PAHs including phenanthrene (C$_{14}$H$_{10}$), chrysene (C$_{18}$H$_{12}$) and benz[a]anthracene (C$_{18}$H$_{12}$); and (iii) compact or pericondensed PAHs including pyrene (C$_{16}$H$_{10}$) and perylene (C$_{20}$H$_{12}$). The symmetries of these molecules are as follows: naphthalene (D$_{2h}$), anthracene(D$_{2h}$), tetracene (D$_{2h}$), phenanthrene (C$_{2v}$), chrysene (C$_{2h}$), benz[a]anthracene (C$_{s}$), pyrene (D$_{2h}$) and perylene (D$_{2h}$). 

\subsection{The importance of anharmonicity}
\begin{figure}
\includegraphics[width=1.0\columnwidth]{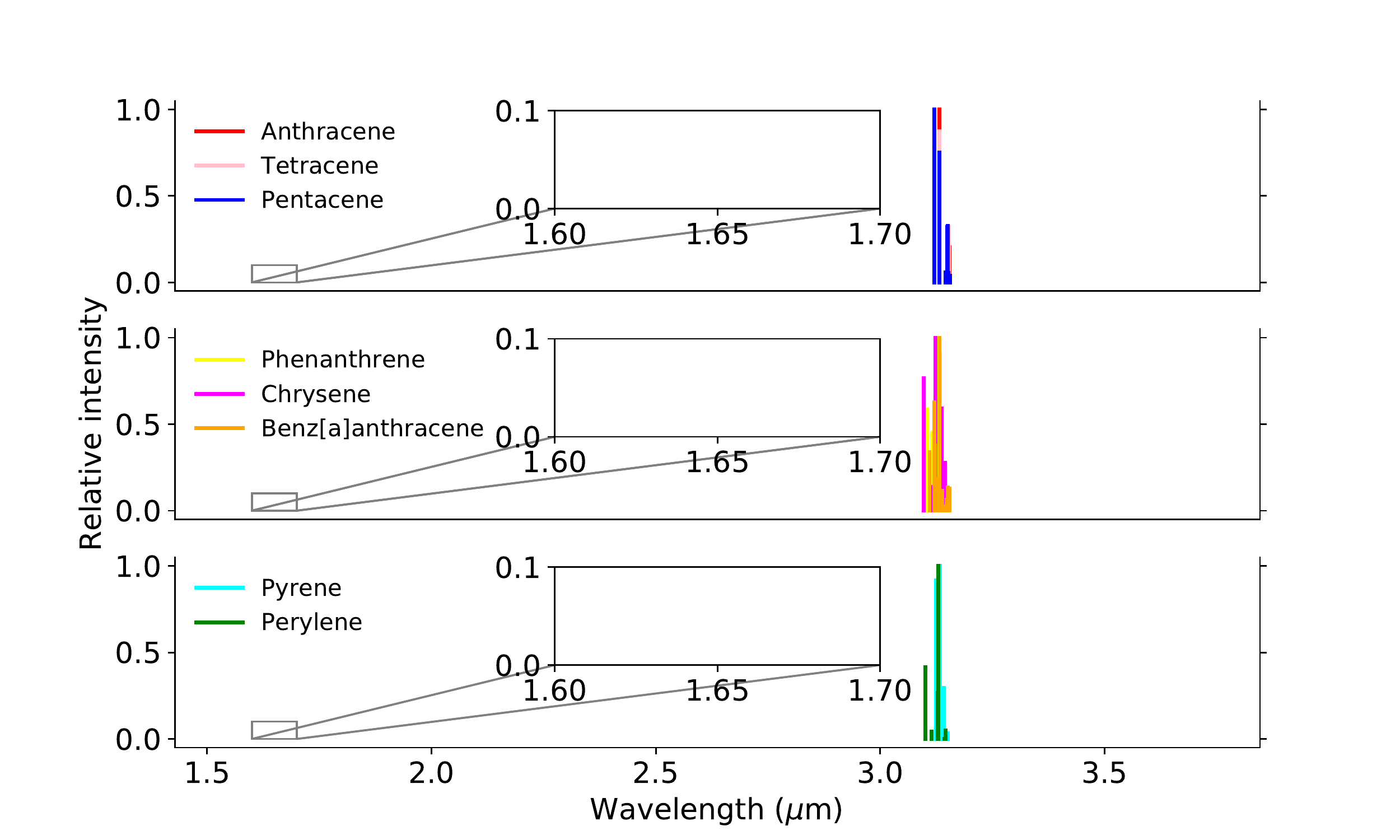}
\caption{Calculated harmonic fundamental bands of anthracene, tetracene, pentacene, phenanthrene, chrysene benz[a]anthracene, pyrene and perylene. The insets zoom in the 1.6--1.7$\mum$ region.}
\label{fig:fundamental}
\end{figure}

For harmonic oscillators, the vibrational states are equally separated on symmetric closed parabolic potential energy surface. The vibrational energies at each state can be calculated from the following equation: 
\begin{equation}
E_{\rm harm}(n) = \sum_i h\nu_i \left(n_i + \frac{1}{2}\right) ~~, 
\label{eq1}
\end{equation}
where $\nu_i$ is the frequency of the $i$-th vibrational mode. {n} $\equiv$ (n$_1$, n$_2$, ...) represents the quantum number of each vibrational state. The fundamental vibrational energy corresponds to the transition from the ground ($n = 0$) to the first vibrational state ($n = 1$), i.e., $E(1) = \left(3/2\right)\sum_i h \nu_i$. The calculated fundamental bands for the studied molecules in the 1.5--3.5$\mum$ wavelength region are shown in Figure~\ref{fig:fundamental}. The insets in Figure~\ref{fig:fundamental} expand the 1.6--1.7$\mum$ region. It is clear that the fundamental bands for PAHs are only present in the C--H stretch region at $\simali$3.1--3.5$\mum$ and no bands are seen in the 1.6--1.7$\mum$ region.

According to Eq.~\ref{eq1}, a molecule bonded by harmonic potential might reach infinite energy $E(n \rightarrow \infty)$ without bond breaking. However, this is not the case in reality. Molecules do dissociate at highly excited vibrationally states \citep{chen2015, chen2019formation}. The potential energy surface is not a symmetric parabolic shape, in contrast, it is a non-symmetric open well (anharmonic potential), which allows a molecule to break at certain vibrationally state. On an anharmonic potential energy surface, the energy levels are unequally separated, which can be computed as follows:
\begin{equation}
E_{\rm anharm}(n) = \sum_i h\nu_i \left(n_i + \frac{1}{2}\right) + \sum_{i \le j}\chi_{ij}\left(n_i + \frac{1}{2}\right)\left(n_j + \frac{1}{2}\right)~~,
\label{eq2}
\end{equation}
where $\chi_{ij}$ represents the anharmonic coupling which describes the interactions (mode couplings) or resonances among various vibrational modes and is usually given by a 2-dimensional matrix. Since most of the elements of $\chi_{ij}$ are negative, the anharmonic energy levels $E_{\rm anharm}(n)$ are lower than the corresponding harmonic energy levels $E_{\rm harm}(n)$. Therefore, with anharmonic potential, the positions and intensities of the fundamental bands differ from that of harmonic potential. Moreover, due to mode couplings, combination bands (i.e., two or more fundamental vibrations are excited simultaneously) show up in the spectra \citep{mackie2015anharmonic,maltseva2015high}. 

\begin{figure}
\includegraphics[width=1.0\columnwidth]{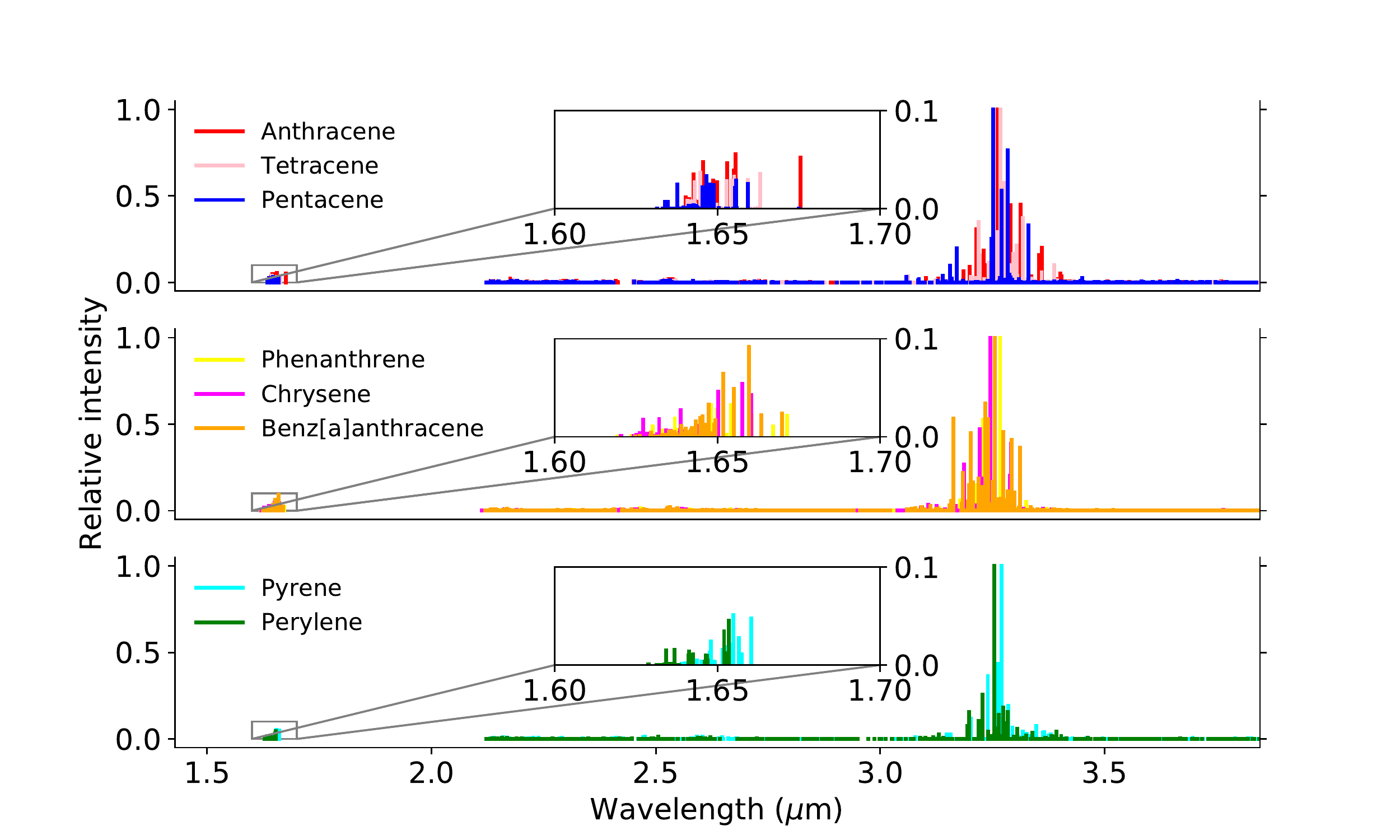}
\caption{Calculated anharmonic IR spectra (including fundamental bands, combination bands and overtones) of anthracene, tetracene, pentacene, phenanthrene, chrysene benz[a]anthracene, pyrene and perylene at ground states. The insets zoom in the 1.6--1.7$\mum$ region.}
\label{fig:anharmonic}
\end{figure}

Figure~\ref{fig:anharmonic} shows the anharmonic spectra of the studied PAHs. Due to anharmonic effects, more bands (e.g., combination bands) appear in the C--H stretch region and the fundamental bands systemically shift to longer wavelengths in comparison with Figure~\ref{fig:fundamental}. The insets in Figure~\ref{fig:anharmonic} amplify the 1.6--1.7$\mum$ region. Unlike Figure~\ref{fig:fundamental}, multiple weak bands show up in this region. The mode numbers, wavelengths and absolute intensities for the C--H stretch and 1.6--1.7$\mum$ regions are given in the Appendix, in addition, the mode descriptions (including frequencies, symmetries, and displacement vectors) for the modes contributing to these two regions can also be found in the Appendix. One can see that the 1.6--1.7$\mum$ region is dominated by the combination bands and no intensive overtone can be found in this region for linear (e.g., anthracene, tetracene and pentacene) and compact (e.g., pyrene and perylene) PAHs, since the first overtones (for the fundamental bands in the C--H stretch region) are electric-dipole-forbidden \citep{duley1994overtone}. 

Only a few overtones can be seen for non-linear PAHs (e.g., phenanthrene, benz[a]anthracene). The highest intensity of the overtones among all the studied molecules is $\simali$2.683$\km\mol^{-1}$ which comes from the 1.66$\mum$ band of benz[a]anthracene, and it corresponds to the overtone of the fundamental band $\nu_{10}$. The second highest overtone comes from phenanthrene with an intensity of $\simali$1.434$\km\mol^{-1}$, which corresponds to the overtone of the fundamental band $\nu_{9}$. 

\subsection{Vibrational excitations}
Molecules in the ISM are mostly excited by starlight (i.e., UV and visible photons). Following the internal conversion, the absorbed energies are rapidly transfered to nuclear degrees of freedom, i.e., vibrational excited states. At a vibrationally excited state, the IR spectrum differs significantly from that of the ground state \citep{chen2018temperature, mackie2018fully}. Moreover, PAHs may dissociate to smaller fragments or isomerize to multiple structures at a high vibrational state, in which anharmonicity plays a key role \citep{chen2019formation}. To incorporate the vibrational excitations, the vibrational microcanonical density of states (DoS) is calculated following the Wang--Landau random walk model \citep{wang2001efficient}. In the model, the initial density of states $g(E)$ is set to 1 for all possible energies $E$. Then, a random walk in energy space is begun by forming trial states, each of which is produced by randomly picking a quanta $n_i$ and randomly changing its value (+1, 0, or -1). In general, if $E_1$ and $E_2$ are energies before and after a quanta set is changed, the transition probability from energy $E_1$ to $E_2$ is
\begin{equation}
p(E_1 \rightarrow E_2) = {\rm min}\left\{\frac{g(E_1)}{g(E_2)},1\right\}~~,
\label{eq3}
\end{equation}
which implies that if $g(E_2) \le g(E_1)$, the state with energy $E_2$ is accepted, otherwise it is accepted with a probability $g(E_1)/g(E_2)$. In addition, the maximum vibrational states have to be considered. All the vibrational quantum number is chosen such that the associated energy remains in the increasing region. 

\begin{figure}
\includegraphics[width=1.0\columnwidth]{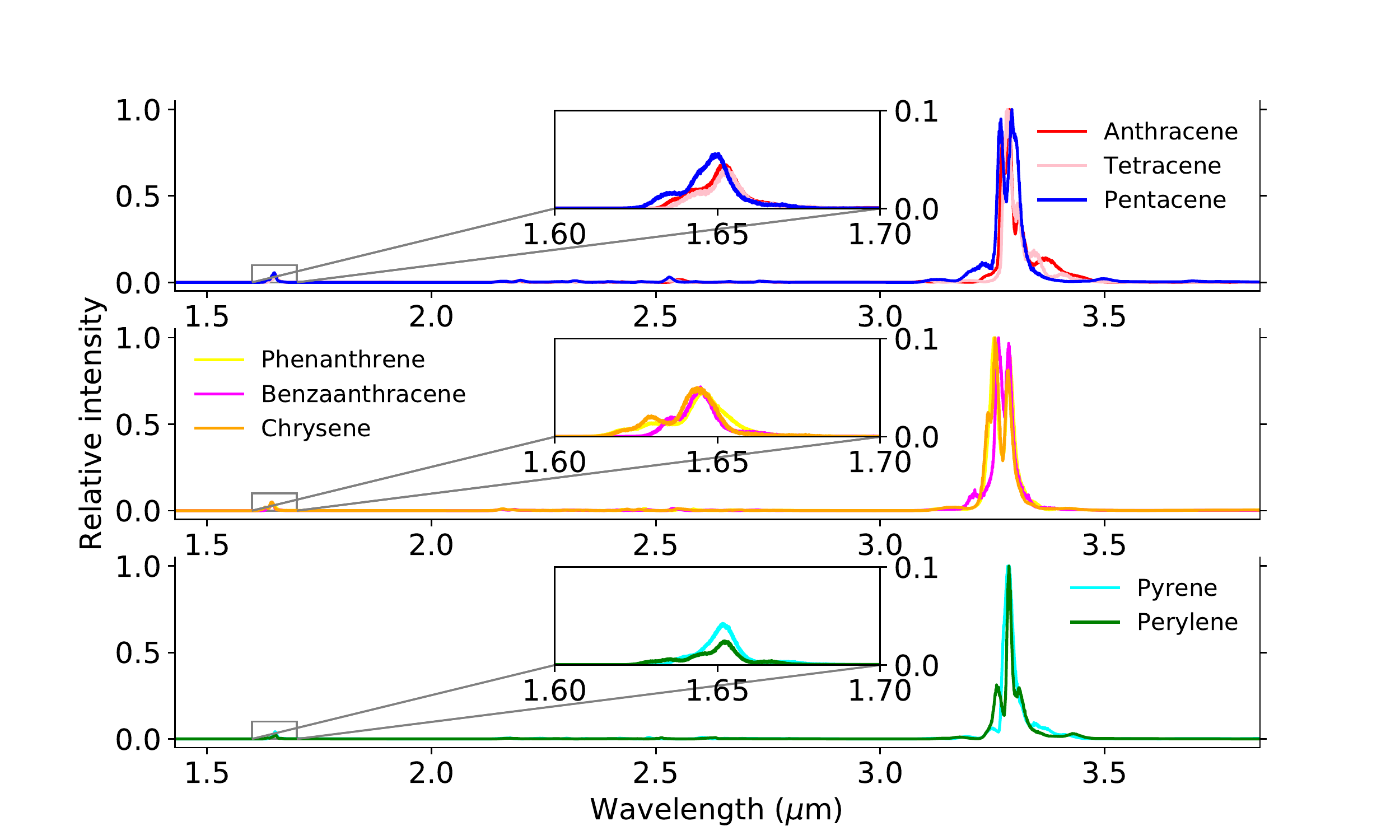}
\caption{Calculated anharmonic IR spectra of anthracene, tetracene, pentacene, phenanthrene, chrysene, benz[a]anthracene, pyrene and perylene at 1000\,K.}
\label{fig:1000K}
\end{figure}

The vibrational DoS provides the weights needed to achieve flat-histogram sampling in energy space \citep{basire2009}. Following the trend of the DoS, second random walks in the space of quantum numbers are performed to build the accumulated absorption $I(\nu, E)$ at the wavelength $\nu$ and for a given internal energy. The absorption intensity $I(\nu, T)$ at finite temperature $T$ can be derived by a standard Laplace transformation of the accumulated absorption $I(\nu, E)$: 
\begin{equation}
I(\nu, T) = \frac{1}{Z}\int I(\nu, E)\,\Omega(E)\,\exp\left(-E/k_B T\right)\,dE ~~,
\label{eq4}
\end{equation}
where $\Omega(E)$ represents the DoS, $k_B$ is the Boltzmann constant, and $Z$ is the partition function: 
\begin{equation}
Z = \int \Omega(E)\,\exp\left(-E/k_B T\right)\,dE ~~,
\label{eq5}
\end{equation}
Figure~\ref{fig:1000K} shows the anharmonic IR spectra of eight molecules at 1000\,K. At such high temperature, the bands are combined to several broad bands. 

\subsection{The band-intensity ratios of $I_{3.1-3.5}/I_{1.6-1.7}$}
As shown in the Appendix, there are a large number of combination bands located in the 1.6--1.7$\mum$ region. However, their average intensities are much lower than the bands in the C--H stretching region. To quantitate the differences between these two regions, we compute, $I_{3.1-3.5}/I_{1.6-1.7}$, the intensity ratios of the C--H stretch region and 1.6--1.7$\mum$ region. In the calculations, the C--H stretch region is measured from 2800--3200$\cm^{-1}$, i.e., 3.125--3.571$\mum$, while the 1.6--1.7$\mum$ region is measured from 5900--6300$\cm^{-1}$, i.e., 1.587--1.695$\mum$. 

Table~\ref{tab:ratio} shows the ratios calculated from the intensities in both regions. At ground vibrational states, pentacene shows the highest ratio ($\simali$13.27). At excited vibrational states, the highest ratio comes from perylene. No general trend as a function of molecular size or structure is found; however, for linear molecules at ground state, the ratio appears to increase with the number of aromatic rings. The average ratio of the studied molecules at ground states are $\langle I_{3.1-3.5}/I_{1.6-1.7}\rangle\approx12.6$. For vibrationally excited states, the average ratio is $\langle I_{3.1-3.5}/I_{1.6-1.7}\rangle\approx17.6$. We notice that the ratios increase $\simali$10--50\% going from ground states to excited states. Hence, for PAHs, the band ratio of 12.6 ought to be a lower limit.

\begin{table*}
\begin{center}
\caption{The intensity ratio, $I_{3.1-3.5}/I_{1.6-1.7}$, of the C--H stretch to the bands in the 1.6--1.7$\mum$ region (with 0\,K for ground states and 1000\,K for vibrational excited states. The ``Max'' column shows the ratios calculated based on the maximum intensities in each region. The ``Sum'' column is calculated based on the total intensities in each region.}
\begin{tabular}{l|cc|cc}
\hline
\hline
Molecules&	Max (0 K) & Sum (0 K) &	Max (1000 K) & Sum (1000 K) \\
\hline
Anthracene&	18.14&	13.08& 22.55&	18.04\\
Tetracene&	27.26&	12.74& 24.63&	18.35 \\
Pentacene&	30.02&	13.27& 17.93&	16.16 \\
Phenanthrene&	31.03&	13.26& 21.95&	15.44 \\
Chrysene&	18.53&	10.84& 19.81&	18.96 \\
Benz[a]anthracene&	10.91&	12.96& 20.05&	13.99 \\
Pyrene&	19.65&	12.98 &23.78&	17.72 \\
Perylene&	22.14&	12.27& 41.30&	21.81 \\
\hline
Mean& 22.19& 12.56& 24.0 &17.56 \\
\hline
\hline
\end{tabular}
\label{tab:ratio}
\end{center}
\end{table*}

\section{Conclusions}
Using DFT in combination with the VPT2 method, the 1.6--1.7$\mum$ region of the IR spectra of PAHs is investigated. The calculations reveal that anharmonicity and mode couplings are crucial for this region, in which combination bands are more intensive and abundant than the overtones. Only two of the eight molecules considered here exhibit overtones in this region (i.e., phenanthrene and benz[a]anthracene, with a maximum intensity of $\simali$2.683$\km\mol^{-1}$). 

To model the vibrational excitations of these molecules, the Wang-Landau random walk algorithm is incorporated. The model shows that, due to anharmonicity and vibrational excitations, the fundamental and combination bands (at the ground state) are merged to several inseparable broad bands. The C--H stretching region at 3.1--3.5$\mum$ is compared to the 1.6--1.7$\mum$ region. The average band intensity ratios are $\langle I_{3.1-3.5}/I_{1.6-1.7}\rangle\approx 12.6$ and $\langle I_{3.1-3.5}/I_{1.6-1.7}\rangle\approx 17.6$, respectively at ground states and at vibrationally excited states. Since the band intensity ratio $I_{3.1-3.5}/I_{1.6-1.7}$ goes up as the molecule gets excited, $\langle I_{3.1-3.5}/I_{1.6-1.7}\rangle\approx 12.6$ should be the lower-limit for PAHs. 

\section*{Acknowledgements}
We thank the anonymous referee for his/her very helpful comments which considerably improved the presentation of this work. This work is supported by the Swedish Research Council (Contract No. 2015--06501). The calculations were performed on resources provided by the Swedish National Infrastructure for Computing (SNIC). A.L. is supported in part by NSF AST-1816411 and NASA 80NSSC19K0572.

\clearpage

\chead{\Large{\textbf{Appendix}}}

\begin{table*}
\begin{center}
\caption{The dominant bands (with intensity $\simgt 0.2\km\mol^{-1}$) in C--H stretching region and 1.6--1.7$\mum$ region for anthracene tetracene and pentacene. The symbol of "+" in the ``Mode'' column represents the combination bands between the modes on both sides of "+".}
\begin{tabular}{lcc|lcc|lcc}
\hline
\hline
\multicolumn{3}{c|}{Anthracene} &\multicolumn{3}{c|}{Tetracene}&\multicolumn{3}{c}{Pentacene}\\
\hline
Mode & Wavelength & Intensity &Mode & Wavelength & Intensity &Mode & Wavelength & Intensity\\
         & ($\mu$m) & (km$\mol^{-1}$) & & ($\mu$m) & (km$\mol^{-1}$) & & ($\mu$m) &  (km$\mol^{-1}$)  
\\		
\hline		
$\nu_{2}$	&	3.251	&	26.494	&	$\nu_{1}$	&	3.269	&	0.263	&	$\nu_{1}$	&	3.249	&	1.447	\\
$\nu_{3}$	&	3.261	&	32.837	&	$\nu_{2}$	&	3.259	&	15.336	&	$\nu_{2}$	&	3.252	&	67.771	\\
$\nu_{6}$	&	3.273	&	17.563	&	$\nu_{3}$	&	3.267	&	52.686	&	$\nu_{3}$	&	3.27	&	35.98	\\
$\nu_{7}$	&	3.306	&	3.858	&	$\nu_{6}$	&	3.276	&	30.334	&	$\nu_{5}$	&	3.275	&	0.207	\\
$\nu_{10}$	&	3.291	&	14.66	&	$\nu_{7}$	&	3.318	&	19.531	&	$\nu_{6}$	&	3.271	&	12.607	\\
$\nu_{14}$+$\nu_{12}$	&	3.13	&	0.699	&	$\nu_{11}$	&	3.293	&	5.278	&	$\nu_{7}$	&	3.265	&	0.412	\\
$\nu_{15}$+$\nu_{13}$	&	3.185	&	2.076	&	$\nu_{16}$+$\nu_{15}$	&	3.132	&	0.409	&	$\nu_{10}$	&	3.331	&	22.409	\\
$\nu_{15}$+$\nu_{14}$	&	3.214	&	10.125	&	$\nu_{17}$+$\nu_{14}$	&	3.146	&	0.483	&	$\nu_{11}$	&	3.287	&	3.057	\\
$\nu_{16}$+$\nu_{12}$	&	3.199	&	2.947	&	$\nu_{17}$+$\nu_{15}$	&	3.156	&	1.075	&	$\nu_{14}$	&	3.284	&	51.692	\\
$\nu_{16}$+$\nu_{15}$	&	3.305	&	5.656	&	$\nu_{18}$+$\nu_{13}$	&	3.137	&	0.702	&	$\nu_{21(2)}$	&	3.252	&	0.233	\\
$\nu_{17}$+$\nu_{11}$	&	3.232	&	3.317	&	$\nu_{18}$+$\nu_{16}$	&	3.202	&	1.597	&	$\nu_{19}$+$\nu_{15}$	&	3.129	&	0.546	\\
$\nu_{17}$+$\nu_{13}$	&	3.287	&	3.695	&	$\nu_{18}$+$\nu_{17}$	&	3.219	&	18.451	&	$\nu_{19}$+$\nu_{17}$	&	3.155	&	6.568	\\
$\nu_{17}$+$\nu_{16}$	&	3.404	&	1.157	&	$\nu_{19}$+$\nu_{16}$	&	3.221	&	8.238	&	$\nu_{20}$+$\nu_{16}$	&	3.139	&	2.544	\\
$\nu_{18}$+$\nu_{11}$	&	3.23	&	6.061	&	$\nu_{19}$+$\nu_{17}$	&	3.242	&	5.966	&	$\nu_{20}$+$\nu_{18}$	&	3.17	&	0.698	\\
$\nu_{18}$+$\nu_{13}$	&	3.269	&	6.153	&	$\nu_{20}$+$\nu_{14}$	&	3.21	&	1.408	&	$\nu_{21}$+$\nu_{15}$	&	3.145	&	0.492	\\
$\nu_{18}$+$\nu_{14}$	&	3.318	&	0.515	&	$\nu_{20}$+$\nu_{16}$	&	3.276	&	1.302	&	$\nu_{21}$+$\nu_{17}$	&	3.171	&	13.395	\\
$\nu_{18}$+$\nu_{16}$	&	3.413	&	0.229	&	$\nu_{20}$+$\nu_{19}$	&	3.338	&	1.148	&	$\nu_{21}$+$\nu_{20}$	&	3.247	&	17.088	\\
$\nu_{19}$+$\nu_{15}$	&	3.394	&	0.669	&	$\nu_{21}$+$\nu_{16}$	&	3.305	&	2.087	&	$\nu_{22}$+$\nu_{15}$	&	3.159	&	1.503	\\
$\nu_{20}$+$\nu_{11}$	&	3.313	&	14.779	&	$\nu_{22}$+$\nu_{13}$	&	3.232	&	0.985	&	$\nu_{22}$+$\nu_{17}$	&	3.184	&	0.706	\\
$\nu_{20}$+$\nu_{13}$	&	3.354	&	5.15	&	$\nu_{22}$+$\nu_{16}$	&	3.288	&	3.521	&	$\nu_{23}$+$\nu_{16}$	&	3.171	&	0.394	\\
$\nu_{21}$+$\nu_{12}$	&	3.337	&	1.166	&	$\nu_{22}$+$\nu_{17}$	&	3.343	&	0.404	&	$\nu_{23}$+$\nu_{19}$	&	3.251	&	0.353	\\
$\nu_{22}$+$\nu_{11}$	&	3.36	&	6.574	&	$\nu_{22}$+$\nu_{20}$	&	3.432	&	0.506	&	$\nu_{23}$+$\nu_{21}$	&	3.277	&	1.085	\\
$\nu_{22}$+$\nu_{13}$	&	3.401	&	1.632	&	$\nu_{23}$+$\nu_{15}$	&	3.309	&	0.435	&	$\nu_{24}$+$\nu_{17}$	&	3.267	&	1.608	\\
$\nu_{22}$+$\nu_{14}$	&	3.428	&	0.205	&	$\nu_{23}$+$\nu_{18}$	&	3.39	&	0.468	&	$\nu_{25}$+$\nu_{16}$	&	3.239	&	0.62	\\
$\nu_{2}$+$\nu_{1}$	&	1.643	&	1.145	&	$\nu_{24}$+$\nu_{13}$	&	3.296	&	8.642	&	$\nu_{25}$+$\nu_{19}$	&	3.329	&	0.275	\\
$\nu_{3}$+$\nu_{1}$	&	1.644	&	0.221	&	$\nu_{24}$+$\nu_{17}$	&	3.394	&	0.677	&	$\nu_{25}$+$\nu_{24}$	&	3.447	&	0.596	\\
$\nu_{4}$+$\nu_{2}$	&	1.643	&	0.249	&	$\nu_{25}$+$\nu_{13}$	&	3.304	&	11.28	&	$\nu_{26}$+$\nu_{16}$	&	3.246	&	0.423	\\
$\nu_{4}$+$\nu_{3}$	&	1.646	&	1.564	&	$\nu_{25}$+$\nu_{16}$	&	3.374	&	0.269	&	$\nu_{27}$+$\nu_{15}$	&	3.291	&	1.864	\\
$\nu_{5}$+$\nu_{2}$	&	1.64	&	0.375	&	$\nu_{25}$+$\nu_{17}$	&	3.403	&	0.296	&	$\nu_{27}$+$\nu_{20}$	&	3.388	&	0.329	\\
$\nu_{6}$+$\nu_{1}$	&	1.641	&	0.317	&	$\nu_{26}$+$\nu_{14}$	&	3.327	&	0.92	&	$\nu_{28}$+$\nu_{16}$	&	3.294	&	0.942	\\
$\nu_{6}$+$\nu_{4}$	&	1.647	&	0.234	&	$\nu_{27}$+$\nu_{14}$	&	3.359	&	2.995	&	$\nu_{30}$+$\nu_{15}$	&	3.309	&	1.364	\\
$\nu_{6}$+$\nu_{5}$	&	1.655	&	0.317	&	$\nu_{27}$+$\nu_{15}$	&	3.388	&	5.201	&	$\nu_{33}$+$\nu_{18}$	&	3.402	&	0.397	\\
$\nu_{7}$+$\nu_{4}$	&	1.649	&	0.94	&	$\nu_{2}$+$\nu_{1}$	&	1.643	&	1.413	&	$\nu_{35}$+$\nu_{18}$	&	3.45	&	1.611	\\
$\nu_{7}$+$\nu_{5}$	&	1.653	&	1.522	&	$\nu_{3}$+$\nu_{1}$	&	1.643	&	0.201	&	$\nu_{38}$+$\nu_{18}$	&	3.477	&	0.239	\\
$\nu_{8}$+$\nu_{3}$	&	1.65	&	0.879	&	$\nu_{4}$+$\nu_{2}$	&	1.643	&	0.216	&	$\nu_{2}$+$\nu_{1}$	&	1.638	&	1.656	\\
$\nu_{8}$+$\nu_{6}$	&	1.655	&	1.81	&	$\nu_{4}$+$\nu_{3}$	&	1.645	&	1.933	&	$\nu_{4}$+$\nu_{3}$	&	1.647	&	2.257	\\
$\nu_{8}$+$\nu_{7}$	&	1.655	&	1.27	&	$\nu_{5}$+$\nu_{2}$	&	1.641	&	0.427	&	$\nu_{5}$+$\nu_{2}$	&	1.635	&	0.477	\\
$\nu_{10}$+$\nu_{9}$	&	1.675	&	1.698	&	$\nu_{6}$+$\nu_{1}$	&	1.642	&	0.4	&	$\nu_{6}$+$\nu_{1}$	&	1.634	&	0.47	\\
	&		&		&	$\nu_{6}$+$\nu_{4}$	&	1.647	&	0.238	&	$\nu_{6}$+$\nu_{4}$	&	1.643	&	0.241	\\
	&		&		&	$\nu_{6}$+$\nu_{5}$	&	1.655	&	0.416	&	$\nu_{6}$+$\nu_{5}$	&	1.645	&	0.502	\\
	&		&		&	$\nu_{7}$+$\nu_{4}$	&	1.648	&	1.001	&	$\nu_{8}$+$\nu_{4}$	&	1.646	&	1.047	\\
	&		&		&	$\nu_{7}$+$\nu_{5}$	&	1.653	&	1.487	&	$\nu_{8}$+$\nu_{5}$	&	1.645	&	1.455	\\
	&		&		&	$\nu_{8}$+$\nu_{3}$	&	1.649	&	1.056	&	$\nu_{9}$+$\nu_{3}$	&	1.648	&	1.182	\\
	&		&		&	$\nu_{8}$+$\nu_{6}$	&	1.655	&	1.704	&	$\nu_{9}$+$\nu_{6}$	&	1.648	&	1.652	\\
	&		&		&	$\nu_{8}$+$\nu_{7}$	&	1.654	&	1.491	&	$\nu_{9}$+$\nu_{8}$	&	1.649	&	1.623	\\
	&		&		&	$\nu_{10}$+$\nu_{9}$	&	1.659	&	1.526	&	$\nu_{10}$+$\nu_{7}$	&	1.655	&	1.419	\\
	&		&		&	$\nu_{12}$+$\nu_{11}$	&	1.663	&	1.85	&	$\nu_{10}$+$\nu_{9}$	&	1.647	&	0.324	\\
	&		&		&		&		&		&	$\nu_{13}$+$\nu_{11}$	&	1.659	&	1.718	\\
	&		&		&		&		&		&	$\nu_{14}$+$\nu_{12}$	&	1.656	&	1.93	\\
\hline
\hline
\end{tabular}
\label{tab:linear}
\end{center}
\end{table*}

\begin{table*}
\begin{center}
\caption{Same as Table~\ref{tab:linear} but for phenanthrene, chrysene and benz[a]anthracene.}
\small
\begin{tabular}{lcc|lcc|lcc}
\hline
\hline
\multicolumn{3}{c|}{Phenanthrene} &\multicolumn{3}{c|}{Chrysene}&\multicolumn{3}{c}{Benz[a]anthracene}\\
\hline
Mode & Wavelength & Intensity &Mode & Wavelength & Intensity &Mode & Wavelength & Intensity\\
         & ($\mu$m) & (km$\mol^{-1}$) & & ($\mu$m) & (km$\mol^{-1}$) & & ($\mu$m) &  (km$\mol^{-1}$)  
\\
\hline				
$\nu_{1}$	&	3.23	&	14.189	&	$\nu_{2}$	&	3.221	&	24.631	&	$\nu_{1}$	&	3.241	&	15.535	\\
$\nu_{2}$	&	3.232	&	20.762	&	$\nu_{6}$	&	3.245	&	52.127	&	$\nu_{2}$	&	3.234	&	9.942	\\
$\nu_{3}$	&	3.253	&	4.707	&	$\nu_{7}$	&	3.273	&	6.729	&	$\nu_{3}$	&	3.263	&	1.973	\\
$\nu_{4}$	&	3.251	&	5.675	&	$\nu_{10}$	&	3.292	&	20.111	&	$\nu_{4}$	&	3.25	&	4.836	\\
$\nu_{5}$	&	3.267	&	44.507	&	$\nu_{11}$	&	3.289	&	10.557	&	$\nu_{5}$	&	3.256	&	29.281	\\
$\nu_{6}$	&	3.279	&	2.974	&	$\nu_{17}$+$\nu_{14}$	&	3.126	&	1.359	&	$\nu_{6}$	&	3.293	&	12.017	\\
$\nu_{7}$	&	3.279	&	0.576	&	$\nu_{17}$+$\nu_{16}$	&	3.144	&	0.413	&	$\nu_{7}$	&	3.282	&	0.934	\\
$\nu_{8}$	&	3.285	&	3.012	&	$\nu_{18}$+$\nu_{14}$	&	3.166	&	0.941	&	$\nu_{8}$	&	3.275	&	1.614	\\
$\nu_{9}$	&	3.289	&	7.916	&	$\nu_{18}$+$\nu_{16}$	&	3.186	&	13.885	&	$\nu_{9}$	&	3.274	&	13.356	\\
$\nu_{10}$	&	3.325	&	2.267	&	$\nu_{19}$+$\nu_{13}$	&	3.168	&	1.419	&	$\nu_{10}$	&	3.284	&	3.289	\\
$\nu_{1}$(2)	&	1.661	&	0.97	&	$\nu_{19}$+$\nu_{15}$	&	3.182	&	1.593	&	$\nu_{11}$	&	3.299	&	3.102	\\
$\nu_{9}$(2)	&	1.648	&	1.434	&	$\nu_{19}$+$\nu_{17}$	&	3.225	&	9.754	&	$\nu_{12}$	&	3.311	&	10.633	\\
$\nu_{10}$(2)	&	1.671	&	0.966	&	$\nu_{19}$+$\nu_{18}$	&	3.279	&	1.3	&	$\nu_{4}$(2)	&	1.652	&	1.881	\\
$\nu_{14}$(2)	&	3.175	&	1.285	&	$\nu_{20}$+$\nu_{13}$	&	3.211	&	1.553	&	$\nu_{5}$(2)	&	1.67	&	0.698	\\
$\nu_{15}$(2)	&	3.277	&	0.481	&	$\nu_{20}$+$\nu_{15}$	&	3.228	&	9.603	&	$\nu_{8}$(2)	&	1.642	&	0.264	\\
$\nu_{14}$+$\nu_{12}$	&	3.13	&	0.724	&	$\nu_{20}$+$\nu_{18}$	&	3.319	&	0.644	&	$\nu_{9}$(2)	&	1.646	&	0.356	\\
$\nu_{14}$+$\nu_{13}$	&	3.139	&	0.251	&	$\nu_{21}$+$\nu_{14}$	&	3.257	&	2.544	&	$\nu_{10}$(2)	&	1.66	&	2.683	\\
$\nu_{15}$+$\nu_{13}$	&	3.178	&	2.586	&	$\nu_{21}$+$\nu_{16}$	&	3.268	&	1.028	&	$\nu_{12}$(2)	&	1.663	&	0.637	\\
$\nu_{15}$+$\nu_{14}$	&	3.23	&	23.48	&	$\nu_{22}$+$\nu_{13}$	&	3.268	&	1.343	&	$\nu_{17}$(2)	&	3.195	&	1.603	\\
$\nu_{16}$+$\nu_{11}$	&	3.189	&	0.39	&	$\nu_{22}$+$\nu_{15}$	&	3.289	&	0.882	&	$\nu_{18}$(2)	&	3.219	&	5.408	\\
$\nu_{16}$+$\nu_{12}$	&	3.203	&	4.167	&	$\nu_{23}$+$\nu_{14}$	&	3.277	&	0.292	&	$\nu_{17}$+$\nu_{14}$	&	3.13	&	0.271	\\
$\nu_{16}$+$\nu_{13}$	&	3.211	&	6.544	&	$\nu_{23}$+$\nu_{16}$	&	3.296	&	0.39	&	$\nu_{17}$+$\nu_{16}$	&	3.158	&	1.64	\\
$\nu_{16}$+$\nu_{15}$	&	3.301	&	0.338	&	$\nu_{23}$+$\nu_{19}$	&	3.383	&	0.422	&	$\nu_{18}$+$\nu_{15}$	&	3.154	&	0.87	\\
$\nu_{17}$+$\nu_{11}$	&	3.241	&	6.494	&	$\nu_{24}$+$\nu_{13}$	&	3.277	&	0.856	&	$\nu_{18}$+$\nu_{16}$	&	3.163	&	15.627	\\
$\nu_{17}$+$\nu_{12}$	&	3.258	&	1.23	&	$\nu_{24}$+$\nu_{15}$	&	3.292	&	0.327	&	$\nu_{18}$+$\nu_{17}$	&	3.202	&	13.144	\\
$\nu_{17}$+$\nu_{13}$	&	3.266	&	2.491	&	$\nu_{25}$+$\nu_{14}$	&	3.339	&	0.224	&	$\nu_{19}$+$\nu_{13}$	&	3.181	&	1.214	\\
$\nu_{18}$+$\nu_{12}$	&	3.27	&	0.299	&	$\nu_{26}$+$\nu_{14}$	&	3.35	&	0.451	&	$\nu_{19}$+$\nu_{14}$	&	3.184	&	6.351	\\
$\nu_{18}$+$\nu_{13}$	&	3.281	&	0.513	&	$\nu_{27}$+$\nu_{15}$	&	3.363	&	0.555	&	$\nu_{19}$+$\nu_{15}$	&	3.201	&	1.128	\\
$\nu_{18}$+$\nu_{15}$	&	3.371	&	0.414	&	$\nu_{28}$+$\nu_{14}$	&	3.381	&	0.211	&	$\nu_{19}$+$\nu_{16}$	&	3.225	&	3.986	\\
$\nu_{19}$+$\nu_{11}$	&	3.281	&	1.852	&	$\nu_{2}$+$\nu_{1}$	&	1.627	&	0.915	&	$\nu_{19}$+$\nu_{18}$	&	3.275	&	0.424	\\
$\nu_{19}$+$\nu_{12}$	&	3.291	&	0.539	&	$\nu_{4}$+$\nu_{3}$	&	1.632	&	0.949	&	$\nu_{20}$+$\nu_{13}$	&	3.198	&	4.267	\\
$\nu_{19}$+$\nu_{13}$	&	3.299	&	0.498	&	$\nu_{6}$+$\nu_{5}$	&	1.639	&	1.415	&	$\nu_{20}$+$\nu_{14}$	&	3.207	&	4.773	\\
$\nu_{20}$+$\nu_{11}$	&	3.273	&	2.146	&	$\nu_{7}$+$\nu_{5}$	&	1.636	&	0.317	&	$\nu_{20}$+$\nu_{15}$	&	3.229	&	1.494	\\
$\nu_{21}$+$\nu_{12}$	&	3.368	&	0.351	&	$\nu_{8}$+$\nu_{6}$	&	1.634	&	0.353	&	$\nu_{20}$+$\nu_{16}$	&	3.246	&	1.054	\\
$\nu_{22}$+$\nu_{12}$	&	3.369	&	0.442	&	$\nu_{8}$+$\nu_{7}$	&	1.65	&	2.4	&	$\nu_{20}$+$\nu_{17}$	&	3.285	&	0.21	\\
$\nu_{3}$+$\nu_{2}$	&	1.63	&	0.463	&	$\nu_{9}$+$\nu_{7}$	&	1.638	&	0.203	&	$\nu_{21}$+$\nu_{13}$	&	3.235	&	18.106	\\
$\nu_{4}$+$\nu_{3}$	&	1.637	&	0.838	&	$\nu_{10}$+$\nu_{8}$	&	1.638	&	0.355	&	$\nu_{21}$+$\nu_{15}$	&	3.269	&	1.989	\\
$\nu_{6}$+$\nu_{3}$	&	1.639	&	0.291	&	$\nu_{10}$+$\nu_{9}$	&	1.66	&	2.23	&	$\nu_{21}$+$\nu_{16}$	&	3.275	&	0.831	\\
$\nu_{6}$+$\nu_{4}$	&	1.633	&	0.303	&	$\nu_{11}$+$\nu_{5}$	&	1.64	&	0.295	&	$\nu_{22}$+$\nu_{14}$	&	3.255	&	0.992	\\
$\nu_{7}$+$\nu_{4}$	&	1.633	&	0.276	&	$\nu_{11}$+$\nu_{8}$	&	1.644	&	0.594	&	$\nu_{22}$+$\nu_{15}$	&	3.273	&	0.263	\\
$\nu_{7}$+$\nu_{6}$	&	1.65	&	1.227	&	$\nu_{12}$+$\nu_{6}$	&	1.639	&	0.236	&	$\nu_{23}$+$\nu_{14}$	&	3.275	&	1.162	\\
$\nu_{8}$+$\nu_{6}$	&	1.644	&	0.28	&	$\nu_{12}$+$\nu_{7}$	&	1.645	&	0.542	&	$\nu_{23}$+$\nu_{15}$	&	3.295	&	0.345	\\
$\nu_{8}$+$\nu_{7}$	&	1.643	&	0.668	&	$\nu_{12}$+$\nu_{10}$	&	1.645	&	0.239	&	$\nu_{23}$+$\nu_{16}$	&	3.309	&	0.518	\\
$\nu_{9}$+$\nu_{4}$	&	1.637	&	0.315	&	$\nu_{12}$+$\nu_{11}$	&	1.658	&	2.814	&	$\nu_{25}$+$\nu_{13}$	&	3.311	&	0.97	\\
$\nu_{9}$+$\nu_{6}$	&	1.645	&	0.539	&		&		&		&	$\nu_{25}$+$\nu_{17}$	&	3.388	&	0.22	\\
$\nu_{9}$+$\nu_{8}$	&	1.654	&	1.434	&		&		&		&	$\nu_{26}$+$\nu_{13}$	&	3.335	&	0.574	\\
$\nu_{10}$+$\nu_{5}$	&	1.667	&	0.48	&		&		&		&	$\nu_{27}$+$\nu_{15}$	&	3.385	&	0.266	\\
$\nu_{10}$+$\nu_{9}$	&	1.649	&	0.575	&		&		&		&	$\nu_{6}$+$\nu_{5}$	&	1.647	&	0.97	\\
	&		&		&		&		&		&	$\nu_{7}$+$\nu_{3}$	&	1.64	&	0.238	\\
	&		&		&		&		&		&	$\nu_{7}$+$\nu_{4}$	&	1.639	&	0.226	\\
	&		&		&		&		&		&	$\nu_{8}$+$\nu_{2}$	&	1.639	&	0.335	\\
	&		&		&		&		&		&	$\nu_{9}$+$\nu_{6}$	&	1.64	&	0.204	\\
	&		&		&		&		&		&	$\nu_{9}$+$\nu_{7}$	&	1.645	&	0.617	\\
	&		&		&		&		&		&	$\nu_{10}$+$\nu_{5}$	&	1.646	&	0.273	\\
	&		&		&		&		&		&	$\nu_{10}$+$\nu_{6}$	&	1.645	&	0.568	\\
	&		&		&		&		&		&	$\nu_{10}$+$\nu_{8}$	&	1.655	&	1.439	\\
	&		&		&		&		&		&	$\nu_{11}$+$\nu_{5}$	&	1.649	&	0.493	\\
	&		&		&		&		&		&	$\nu_{11}$+$\nu_{9}$	&	1.649	&	0.368	\\
	&		&		&		&		&		&	$\nu_{11}$+$\nu_{10}$	&	1.643	&	0.457	\\
\hline
\hline
\end{tabular}
\end{center}
\label{tab:non-linear}
\end{table*}

\begin{table*}
\begin{center}
\caption{Same as Table~\ref{tab:linear} but for pyrene and perylene.}
\begin{tabular}{lcc|lcc}
\hline
\hline
\multicolumn{3}{c|}{Pyrene} &\multicolumn{3}{c}{Perylene}\\
\hline
Mode & Wavelength & Intensity &Mode & Wavelength & Intensity \\
         & ($\mu$m) & (km$\mol^{-1}$) & & ($\mu$m) & (km$\mol^{-1}$)
\\
\hline
$\nu_{2}$	&	3.268	&	17.11	&	$\nu_{2}$	&	3.227	&	19.342	\\
$\nu_{4}$	&	3.264	&	10.722	&	$\nu_{3}$	&	3.24	&	3.259	\\
$\nu_{6}$	&	3.271	&	51.48	&	$\nu_{6}$	&	3.284	&	11.855	\\
$\nu_{7}$	&	3.261	&	22.431	&	$\nu_{7}$	&	3.254	&	75.684	\\
$\nu_{10}$	&	3.295	&	3.386	&	$\nu_{10}$	&	3.259	&	4.774	\\
$\nu_{14}$+$\nu_{13}$	&	3.129	&	0.366	&	$\nu_{16}$+$\nu_{15}$	&	3.131	&	2.005	\\
$\nu_{15}$+$\nu_{12}$	&	3.148	&	1.3	&	$\nu_{18}$+$\nu_{14}$	&	3.141	&	0.472	\\
$\nu_{15}$+$\nu_{13}$	&	3.157	&	1.312	&	$\nu_{19}$+$\nu_{15}$	&	3.194	&	5.775	\\
$\nu_{16}$+$\nu_{12}$	&	3.202	&	6.022	&	$\nu_{19}$+$\nu_{17}$	&	3.198	&	11.821	\\
$\nu_{16}$+$\nu_{13}$	&	3.219	&	2.167	&	$\nu_{20}$+$\nu_{13}$	&	3.194	&	0.703	\\
$\nu_{17}$+$\nu_{11}$	&	3.2	&	2.747	&	$\nu_{20}$+$\nu_{16}$	&	3.22	&	7.837	\\
$\nu_{17}$+$\nu_{14}$	&	3.24	&	18.711	&	$\nu_{20}$+$\nu_{18}$	&	3.253	&	1.574	\\
$\nu_{17}$+$\nu_{15}$	&	3.285	&	9.817	&	$\nu_{20}$+$\nu_{19}$	&	3.305	&	4.553	\\
$\nu_{17}$+$\nu_{16}$	&	3.362	&	1.78	&	$\nu_{21}$+$\nu_{13}$	&	3.221	&	1.25	\\
$\nu_{18}$+$\nu_{14}$	&	3.28	&	2.257	&	$\nu_{21}$+$\nu_{16}$	&	3.258	&	1.841	\\
$\nu_{18}$+$\nu_{15}$	&	3.311	&	0.424	&	$\nu_{21}$+$\nu_{18}$	&	3.3	&	0.607	\\
$\nu_{18}$+$\nu_{16}$	&	3.381	&	1.424	&	$\nu_{21}$+$\nu_{19}$	&	3.338	&	2.793	\\
$\nu_{19}$+$\nu_{11}$	&	3.255	&	9.365	&	$\nu_{22}$+$\nu_{14}$	&	3.274	&	13.828	\\
$\nu_{19}$+$\nu_{14}$	&	3.318	&	1.12	&	$\nu_{22}$+$\nu_{15}$	&	3.264	&	10.669	\\
$\nu_{19}$+$\nu_{15}$	&	3.347	&	0.626	&	$\nu_{22}$+$\nu_{17}$	&	3.28	&	6.801	\\
$\nu_{20}$+$\nu_{11}$	&	3.253	&	0.609	&	$\nu_{22}$+$\nu_{20}$	&	3.384	&	0.975	\\
$\nu_{20}$+$\nu_{14}$	&	3.308	&	0.595	&	$\nu_{22}$+$\nu_{21}$	&	3.415	&	0.277	\\
$\nu_{20}$+$\nu_{15}$	&	3.348	&	3.847	&	$\nu_{23}$+$\nu_{13}$	&	3.246	&	1.278	\\
$\nu_{21}$+$\nu_{12}$	&	3.331	&	0.792	&	$\nu_{23}$+$\nu_{16}$	&	3.321	&	0.467	\\
$\nu_{21}$+$\nu_{13}$	&	3.327	&	1.585	&	$\nu_{23}$+$\nu_{19}$	&	3.367	&	0.616	\\
$\nu_{22}$+$\nu_{12}$	&	3.318	&	2.181	&	$\nu_{24}$+$\nu_{13}$	&	3.259	&	0.575	\\
$\nu_{23}$+$\nu_{13}$	&	3.365	&	1.941	&	$\nu_{24}$+$\nu_{14}$	&	3.271	&	1.358	\\
$\nu_{24}$+$\nu_{11}$	&	3.375	&	0.965	&	$\nu_{24}$+$\nu_{17}$	&	3.297	&	1.098	\\
$\nu_{24}$+$\nu_{14}$	&	3.428	&	0.241	&	$\nu_{24}$+$\nu_{20}$	&	3.403	&	1.177	\\
$\nu_{2}$+$\nu_{1}$	&	1.648	&	1.254	&	$\nu_{25}$+$\nu_{13}$	&	3.325	&	1.371	\\
$\nu_{4}$+$\nu_{3}$	&	1.652	&	0.81	&	$\nu_{25}$+$\nu_{18}$	&	3.393	&	3.126	\\
$\nu_{5}$+$\nu_{2}$	&	1.641	&	0.211	&	$\nu_{26}$+$\nu_{15}$	&	3.368	&	0.525	\\
$\nu_{6}$+$\nu_{5}$	&	1.657	&	1.409	&	$\nu_{27}$+$\nu_{16}$	&	3.383	&	1.225	\\
$\nu_{7}$+$\nu_{3}$	&	1.644	&	0.24	&	$\nu_{30}$+$\nu_{13}$	&	3.382	&	1.195	\\
$\nu_{7}$+$\nu_{5}$	&	1.653	&	0.952	&	$\nu_{31}$+$\nu_{17}$	&	3.463	&	0.485	\\
$\nu_{8}$+$\nu_{6}$	&	1.654	&	1.062	&	$\nu_{2}$+$\nu_{1}$	&	1.634	&	1.127	\\
$\nu_{8}$+$\nu_{7}$	&	1.655	&	2.62	&	$\nu_{4}$+$\nu_{3}$	&	1.637	&	1.191	\\
$\nu_{9}$+$\nu_{4}$	&	1.657	&	0.555	&	$\nu_{6}$+$\nu_{5}$	&	1.643	&	0.836	\\
$\nu_{10}$+$\nu_{3}$	&	1.655	&	0.44	&	$\nu_{7}$+$\nu_{5}$	&	1.641	&	0.723	\\
$\nu_{10}$+$\nu_{8}$	&	1.648	&	0.642	&	$\nu_{8}$+$\nu_{6}$	&	1.642	&	0.78	\\
$\nu_{10}$+$\nu_{9}$	&	1.66	&	2.441	&	$\nu_{8}$+$\nu_{7}$	&	1.641	&	1.042	\\
	&		&		&	$\nu_{9}$+$\nu_{6}$	&	1.647	&	0.681	\\
	&		&		&	$\nu_{9}$+$\nu_{7}$	&	1.646	&	0.256	\\
	&		&		&	$\nu_{10}$+$\nu_{8}$	&	1.646	&	0.763	\\
	&		&		&	$\nu_{10}$+$\nu_{9}$	&	1.652	&	2.587	\\
	&		&		&	$\nu_{11}$+$\nu_{5}$	&	1.647	&	0.333	\\
	&		&		&	$\nu_{11}$+$\nu_{8}$	&	1.646	&	0.252	\\
	&		&		&	$\nu_{11}$+$\nu_{9}$	&	1.653	&	0.767	\\
	&		&		&	$\nu_{12}$+$\nu_{7}$	&	1.646	&	0.349	\\
	&		&		&	$\nu_{12}$+$\nu_{10}$	&	1.652	&	0.935	\\
	&		&		&	$\nu_{12}$+$\nu_{11}$	&	1.653	&	3.418	\\
\hline
\hline
\end{tabular}
\end{center}
\label{tab:compact}
\end{table*}

\begin{figure*}
\includegraphics[width=1.0\textwidth]{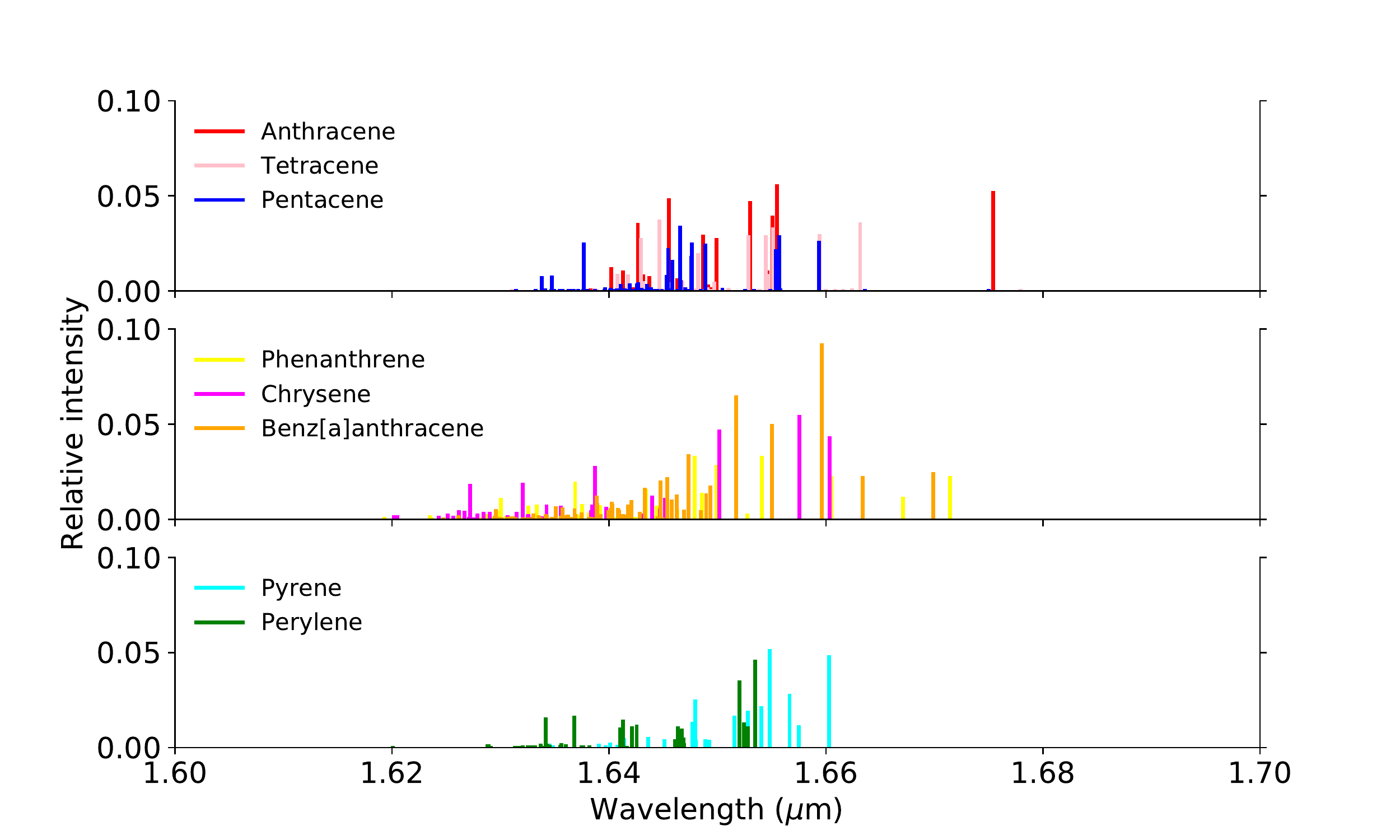}
\caption{The 1.6--1.7$\mum$ region of the anharmonic IR spectra of ground-state anthracene, tetracene, pentacene, phenanthrene, chrysene benz[a]anthracene, pyrene and perylene.}
\label{fig:zoomin_anharmonic}
\end{figure*}

\begin{figure*}
\includegraphics[width=1.0\textwidth]{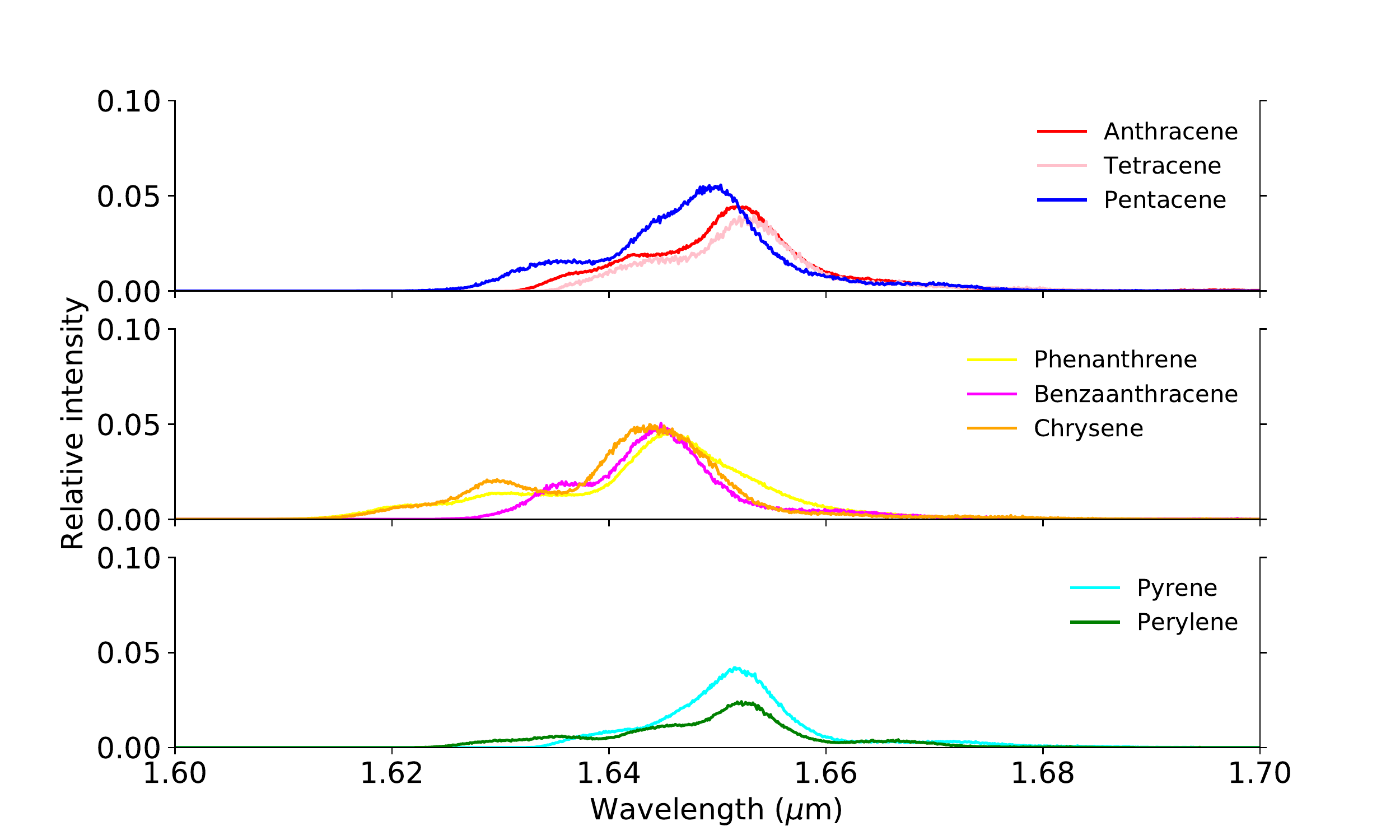}
\caption{The 1.6--1.7$\mum$ region of the vibrationally-excited spectra of anthracene, tetracene, pentacene, phenanthrene, chrysene, benz[a]anthracene, pyrene and perylene at 1000\,K.}
\label{fig:zoomin_1000K}
\end{figure*}

\begin{figure*}
\hspace{-4cm}\includegraphics[width=1.6\textwidth]{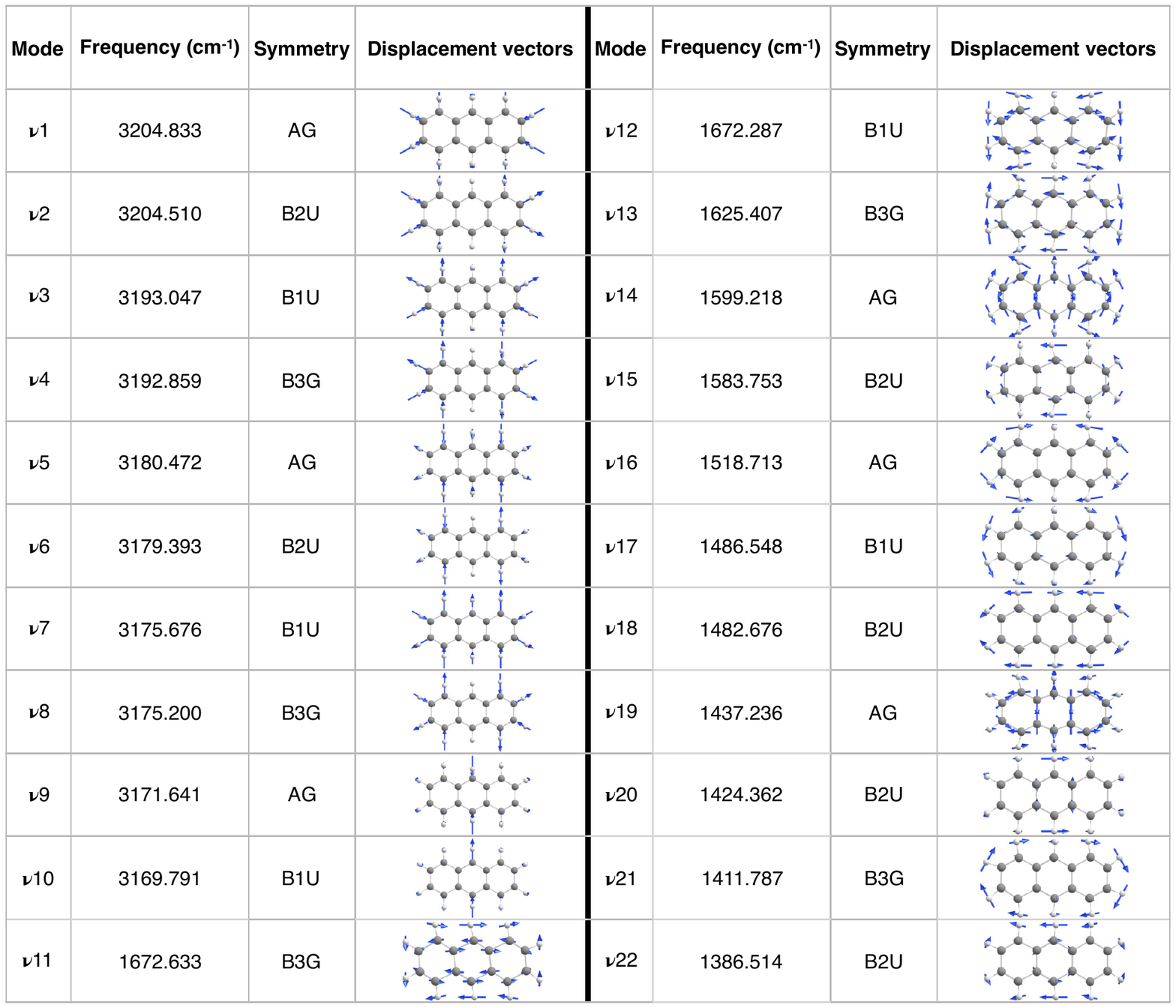}
\caption{(\textbf{Anthracene}) Harmonic frequencies, symmetries and displacement vectors for the modes which contribute to the combination bands or overtones in the 1.6--1.7$\mum$ and C--H stretching regions.}
\label{fig:anthracene_modes}
\end{figure*}

\begin{figure*}
\hspace{-4cm}\includegraphics[width=1.6\textwidth]{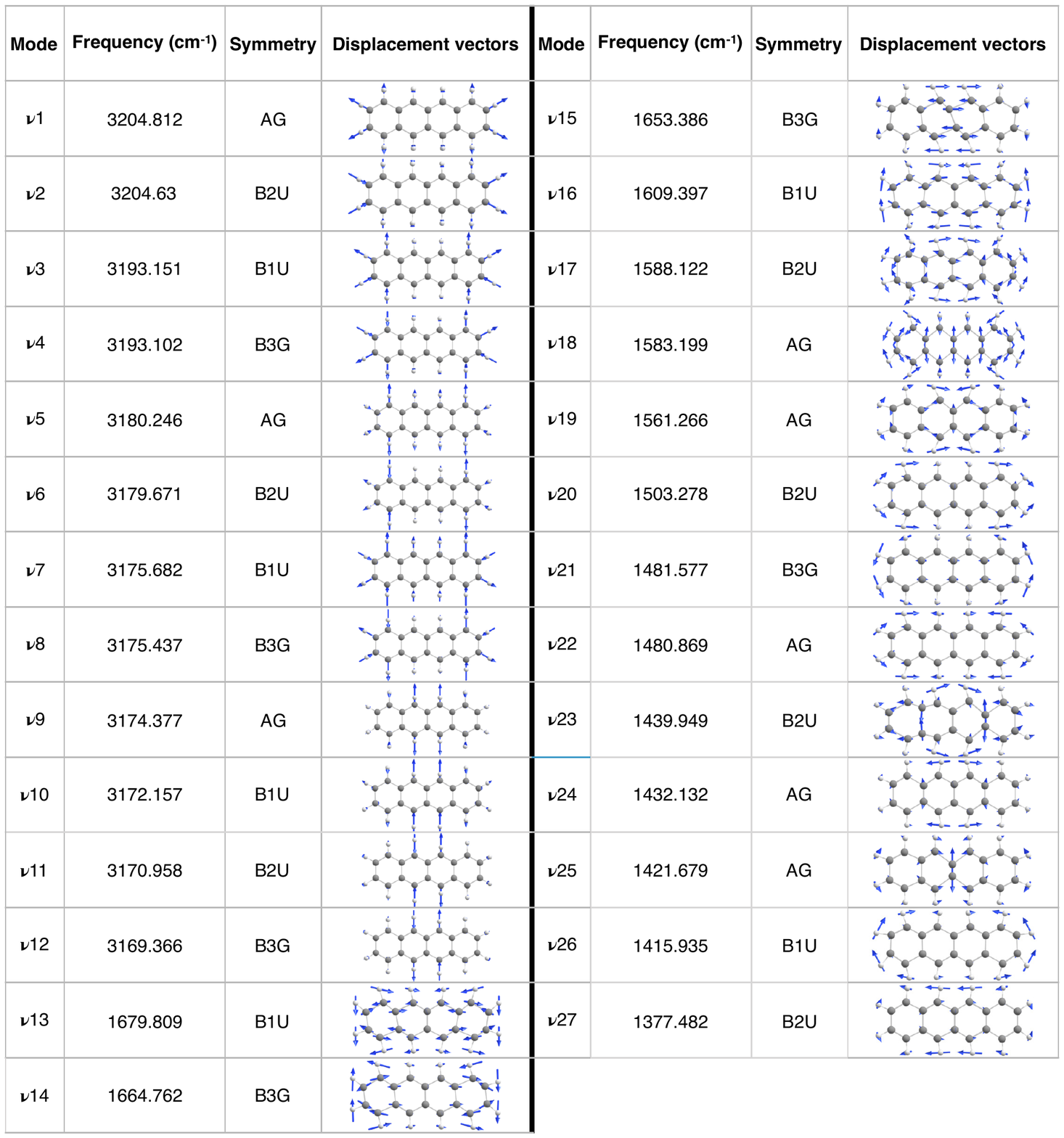}
\caption{(\textbf{Tetracene}) Harmonic frequencies, symmetries and displacement vectors for the modes which contribute to the combination bands or overtones in the 1.6--1.7$\mum$ and C--H stretching regions. }
\label{fig:tetracene_modes}
\end{figure*}

\begin{figure*}
\hspace{-4cm}\includegraphics[width=1.6\textwidth]{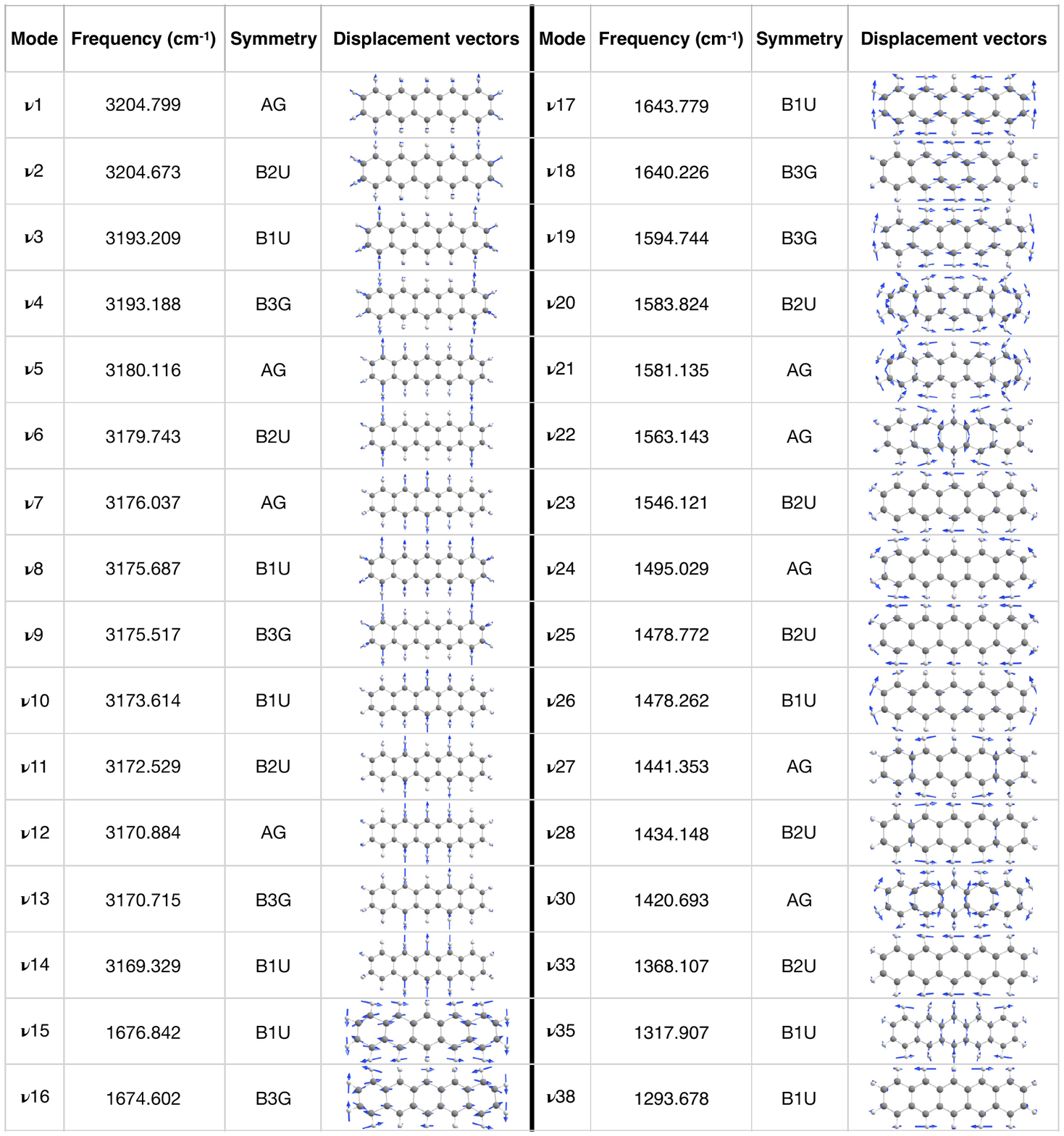}
\caption{(\textbf{Pentacene}) Harmonic frequencies, symmetries and displacement vectors for the modes which contribute to the combination bands or overtones in the 1.6--1.7$\mum$ and C--H stretching regions. }
\label{fig:pentacene_modes}
\end{figure*}

\begin{figure*}
\hspace{-4cm}\includegraphics[width=1.6\textwidth]{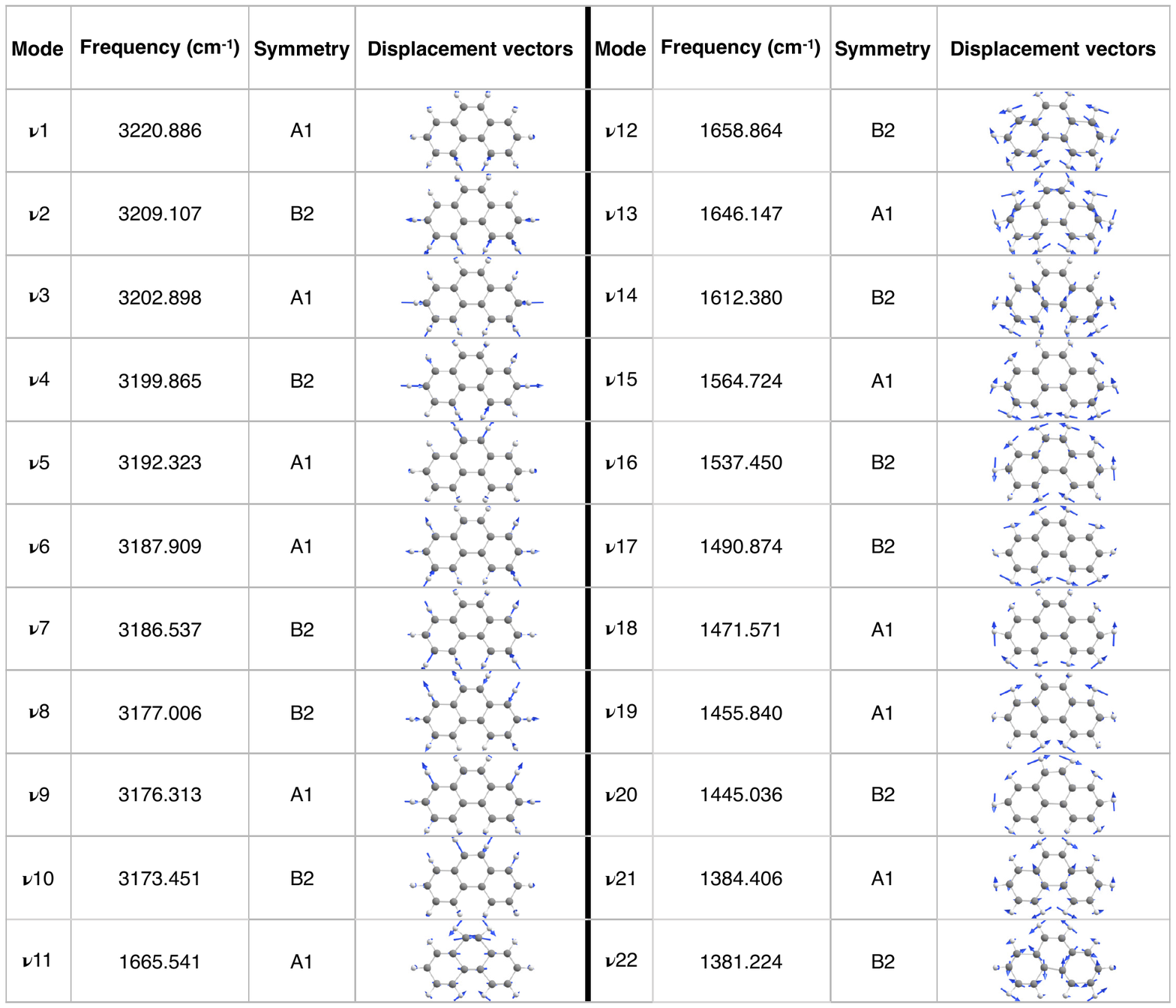}
\caption{(\textbf{Phenanthrene}) Harmonic frequencies, symmetries and displacement vectors for the modes which contribute to the combination bands or overtones in the 1.6--1.7$\mum$ and C--H stretching regions. }
\label{fig:phenanthrene_modes}
\end{figure*}

\begin{figure*}
\hspace{-4cm}\includegraphics[width=1.6\textwidth]{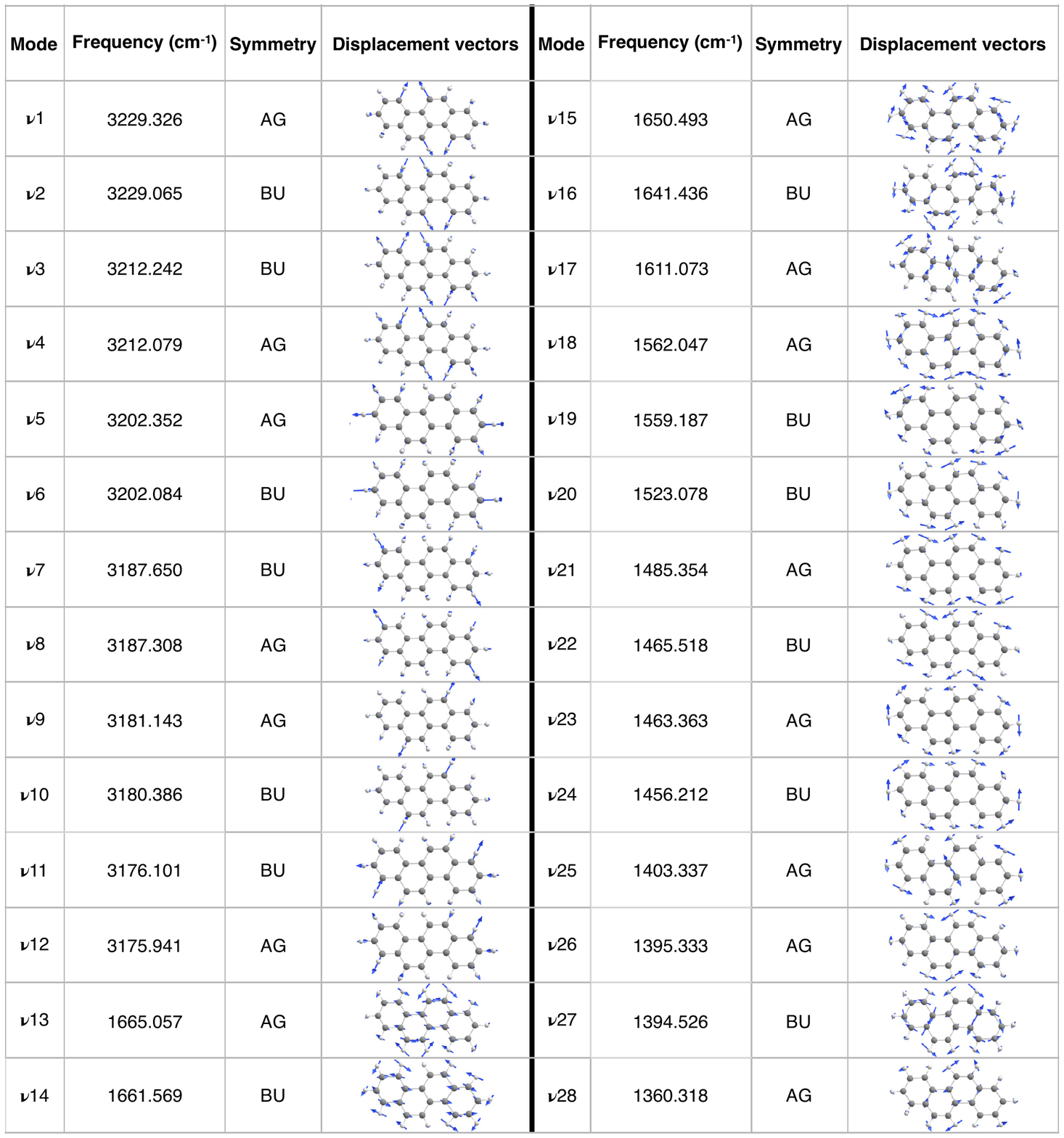}
\caption{(\textbf{Chrysene}) Harmonic frequencies, symmetries and displacement vectors for the modes which contribute to the combination bands or overtones in the 1.6--1.7$\mum$ and C--H stretching regions. }
\label{fig:chrysene_modes}
\end{figure*}

\begin{figure*}
\hspace{-4cm}\includegraphics[width=1.6\textwidth]{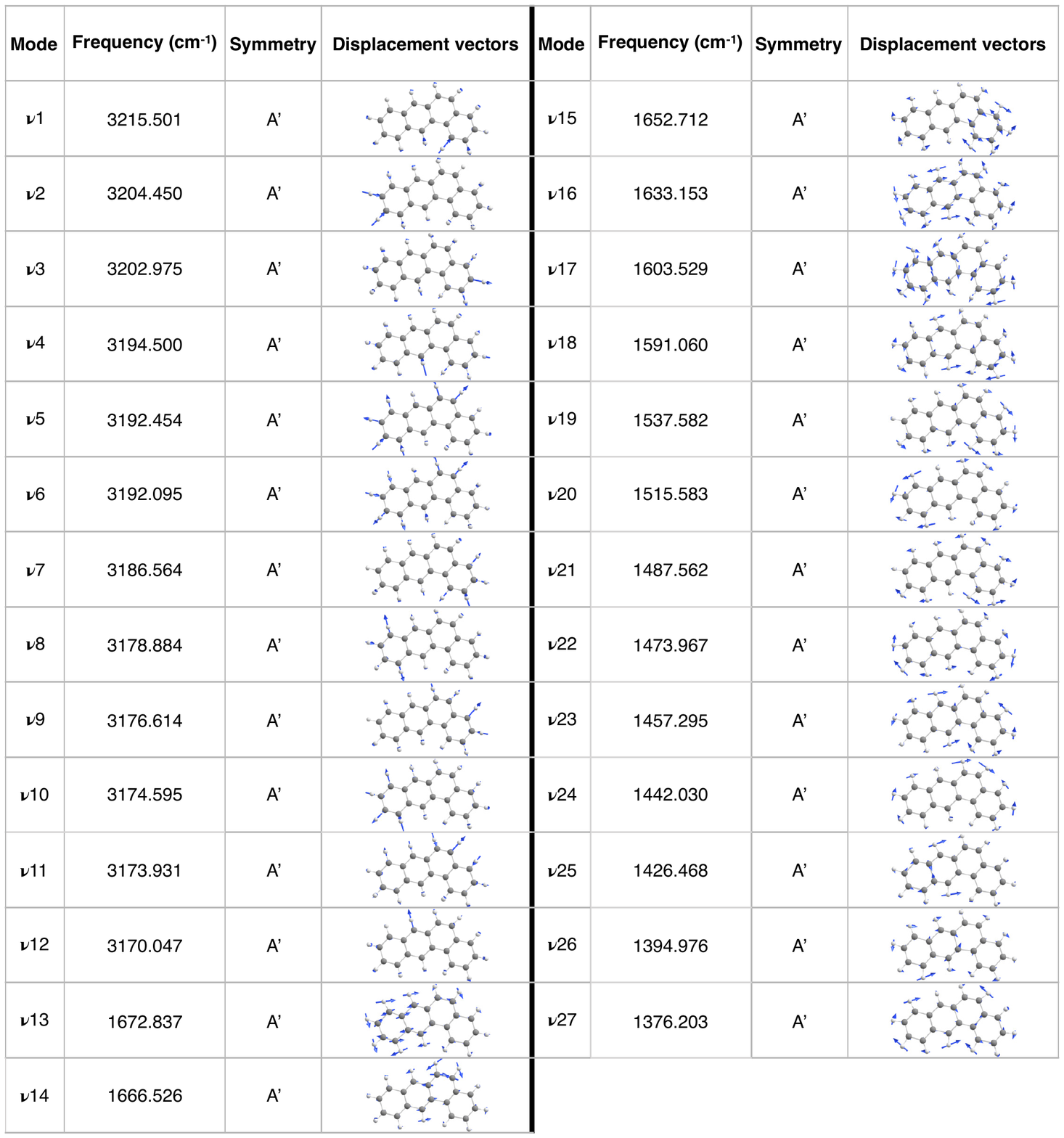}
\caption{(\textbf{Benz[a]anthracene}) Harmonic frequencies, symmetries and displacement vectors for the modes which contribute to the combination bands or overtones in the 1.6--1.7$\mum$ and C--H stretching regions. }
\label{fig:benzaanthracene_modes}
\end{figure*}

\begin{figure*}
\hspace{-4cm}\includegraphics[width=1.6\textwidth]{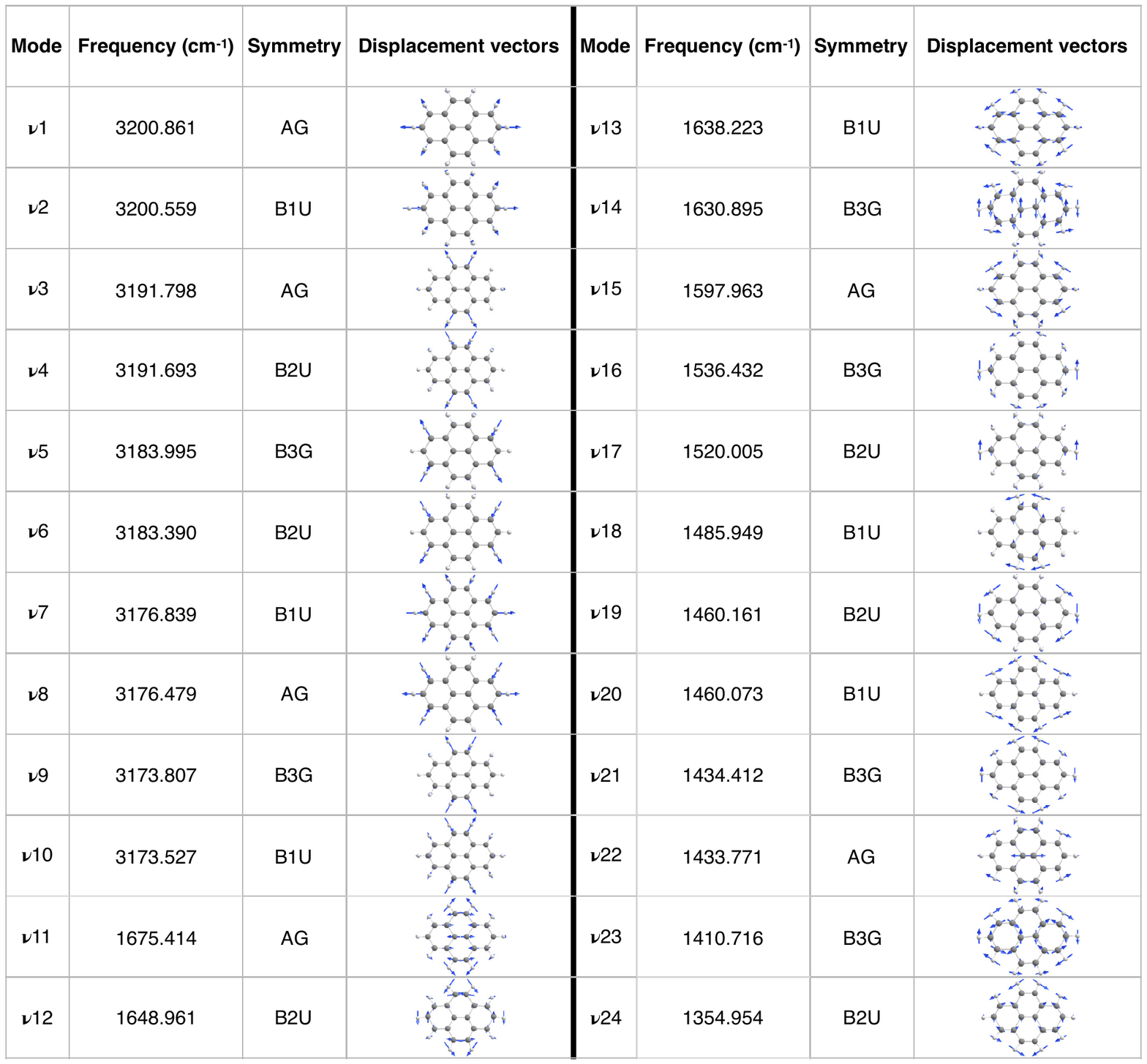}
\caption{(\textbf{Pyrene}) Harmonic frequencies, symmetries and displacement vectors for the modes which contribute to the combination bands or overtones in the 1.6--1.7$\mum$ and C--H stretching regions. }
\label{fig:pyrene_modes}
\end{figure*}

\begin{figure*}
\hspace{-4cm}\includegraphics[width=1.6\textwidth]{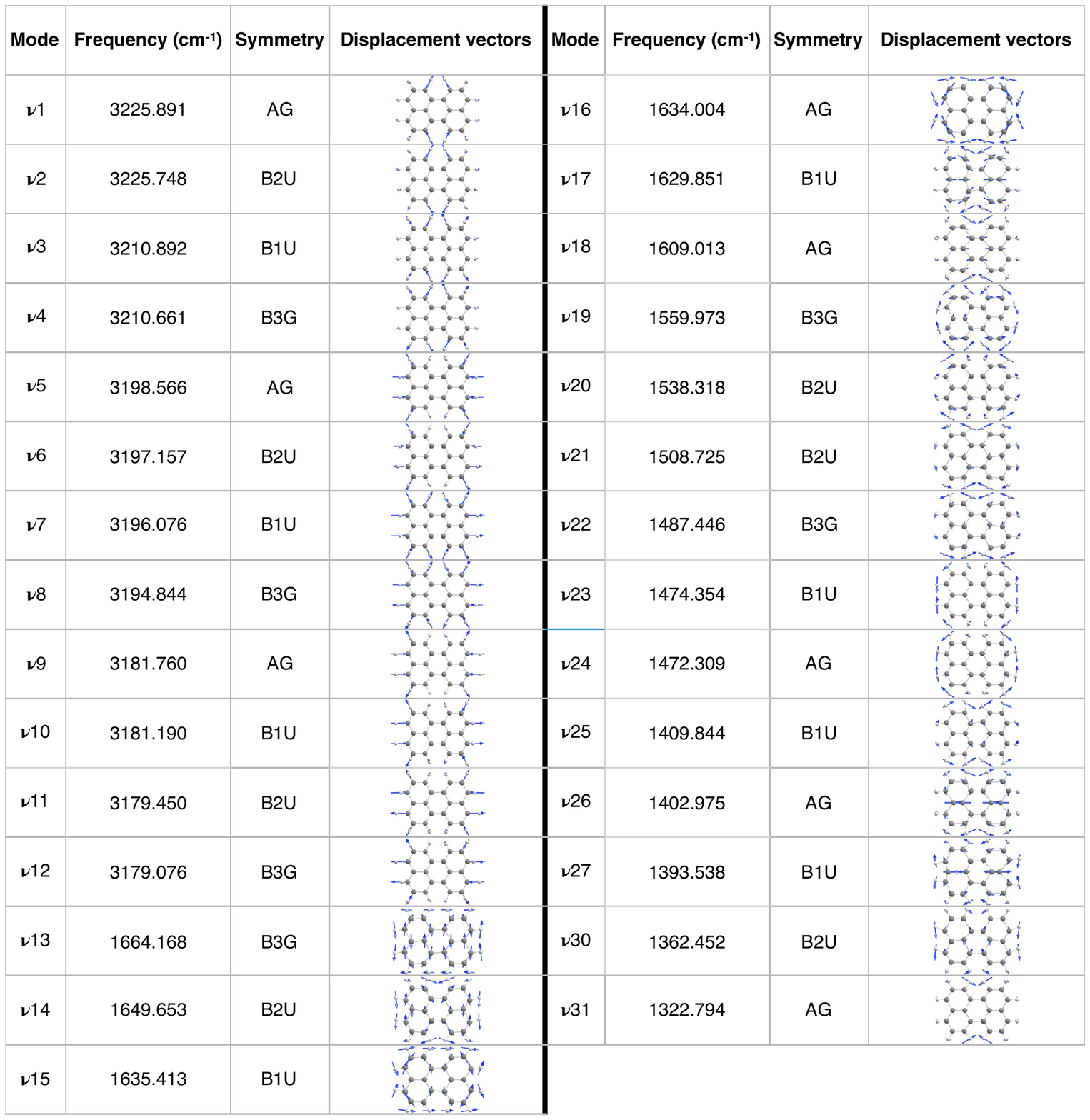}
\caption{(\textbf{Perylene}) Harmonic frequencies, symmetries and displacement vectors for the modes which contribute to the combination bands or overtones in the 1.6--1.7$\mum$ and C--H stretching regions. }
\label{fig:perylene_modes}
\end{figure*}

\end{document}